\documentclass[manuscript]{aastex61}

\usepackage{hyperref}
\usepackage{bm}
\usepackage{amsmath}

\hypersetup{pdftitle= {Improved Empirical Models for Type Ia Supernovae}
pdfauthor = {Clare Saunders},
colorlinks=true,        
linkcolor=blue,         
citecolor=blue,         
urlcolor=blue           
}

\begin{document}
\title{\textsc{SNEMO}: Improved Empirical Models for Type Ia Supernovae}
\shorttitle{SNEMO}

 \author{    C.~Saunders}
\affiliation{ Sorbonne Universit\'e, Universit\'e Paris Diderot, CNRS/IN2P3, Laboratoire de Physique Nucl\'eaire et de Hautes \'Energies, 4 Place Jussieu, Paris, France}
\affiliation{    Sorbonne Universit\'es, Institut Lagrange de Paris (ILP), 98 bis Boulevard Arago, 75014 Paris, France}
\affiliation{    Physics Division, Lawrence Berkeley National Laboratory,
    1 Cyclotron Road, Berkeley, CA, 94720}
    
\author{     G.~Aldering}
\affiliation{    Physics Division, Lawrence Berkeley National Laboratory,
    1 Cyclotron Road, Berkeley, CA, 94720}

\author{     P.~Antilogus}
\affiliation{ Sorbonne Universit\'e, Universit\'e Paris Diderot, CNRS/IN2P3, Laboratoire de Physique Nucl\'eaire et de Hautes \'Energies, 4 Place Jussieu, Paris, France}

\author{     S.~Bailey}
\affiliation{    Physics Division, Lawrence Berkeley National Laboratory,
    1 Cyclotron Road, Berkeley, CA, 94720}

\author{     C.~Baltay}
\affiliation{    Department of Physics, Yale University,
    New Haven, CT, 06250-8121}

\author{     K.~Barbary}
\affiliation{
    Department of Physics, University of California Berkeley,
    366 LeConte Hall MC 7300, Berkeley, CA, 94720-7300}

\author{    D.~Baugh}
\affiliation{   Tsinghua Center for Astrophysics, Tsinghua University, Beijing 100084, China }

\author{     K.~Boone}
\affiliation{    Physics Division, Lawrence Berkeley National Laboratory,
    1 Cyclotron Road, Berkeley, CA, 94720}
\affiliation{
    Department of Physics, University of California Berkeley,
    366 LeConte Hall MC 7300, Berkeley, CA, 94720-7300}

\author{     S.~Bongard}
\affiliation{ Sorbonne Universit\'e, Universit\'e Paris Diderot, CNRS/IN2P3, Laboratoire de Physique Nucl\'eaire et de Hautes \'Energies, 4 Place Jussieu, Paris, France}

\author{     C.~Buton}
\affiliation{    Universit\'e de Lyon, F-69622, Lyon, France ; Universit\'e de Lyon 1, Villeurbanne ;
    CNRS/IN2P3, Institut de Physique Nucl\'eaire de Lyon}

\author{     J.~Chen}
\affiliation{   Tsinghua Center for Astrophysics, Tsinghua University, Beijing 100084, China }

\author{     N.~Chotard}
\affiliation{    Universit\'e de Lyon, F-69622, Lyon, France ; Universit\'e de Lyon 1, Villeurbanne ;
    CNRS/IN2P3, Institut de Physique Nucl\'eaire de Lyon}

\author{     Y.~Copin}
\affiliation{    Universit\'e de Lyon, F-69622, Lyon, France ; Universit\'e de Lyon 1, Villeurbanne ;
    CNRS/IN2P3, Institut de Physique Nucl\'eaire de Lyon}

\author{     S.~Dixon}
\affiliation{    Physics Division, Lawrence Berkeley National Laboratory,
    1 Cyclotron Road, Berkeley, CA, 94720}
\affiliation{
    Department of Physics, University of California Berkeley,
    366 LeConte Hall MC 7300, Berkeley, CA, 94720-7300}

\author{     P.~Fagrelius}
\affiliation{    Physics Division, Lawrence Berkeley National Laboratory,
    1 Cyclotron Road, Berkeley, CA, 94720}
\affiliation{
    Department of Physics, University of California Berkeley,
    366 LeConte Hall MC 7300, Berkeley, CA, 94720-7300}

\author{     H.~K.~Fakhouri}
\affiliation{    Physics Division, Lawrence Berkeley National Laboratory,
    1 Cyclotron Road, Berkeley, CA, 94720}
  \affiliation{
    Department of Physics, University of California Berkeley,
    366 LeConte Hall MC 7300, Berkeley, CA, 94720-7300}

\author{     U.~Feindt}
\affiliation{The Oskar Klein Centre, Department of Physics, AlbaNova, Stockholm University, SE-106 91 Stockholm, Sweden}

\author{     D.~Fouchez}
\affiliation{    Centre de Physique des Particules de Marseille,
    Aix-Marseille Universit\'e , CNRS/IN2P3,
    163 avenue de Luminy - Case 902 - 13288 Marseille Cedex 09, France}

\author{     E.~Gangler}
\affiliation{    Clermont Universit\'e, Universit\'e Blaise Pascal, CNRS/IN2P3, Laboratoire de Physique Corpusculaire,
    BP 10448, F-63000 Clermont-Ferrand, France}

\author{     B.~Hayden}
\affiliation{    Physics Division, Lawrence Berkeley National Laboratory,
    1 Cyclotron Road, Berkeley, CA, 94720}

\author{     W.~Hillebrandt}
\affiliation{    Max-Planck-Institut f\"ur Astrophysik, Karl-Schwarzschild-Str. 1,
D-85748 Garching, Germany}

\author{A.~G.~Kim}
\affiliation{    Physics Division, Lawrence Berkeley National Laboratory,
    1 Cyclotron Road, Berkeley, CA, 94720}

\author{     M.~Kowalski}
\affiliation{    Institut fur Physik,  Humboldt-Universitat zu Berlin,
    Newtonstr. 15, 12489 Berlin}
\affiliation{ DESY, D-15735 Zeuthen, Germany}

\author{     D.~K\"usters}
\affiliation{    Institut fur Physik,  Humboldt-Universitat zu Berlin,
    Newtonstr. 15, 12489 Berlin}

\author{     P.-F.~Leget}
\affiliation{    Clermont Universit\'e, Universit\'e Blaise Pascal, CNRS/IN2P3, Laboratoire de Physique Corpusculaire,
    BP 10448, F-63000 Clermont-Ferrand, France}

\author{     S.~Lombardo}
\affiliation{    Institut fur Physik,  Humboldt-Universitat zu Berlin,
    Newtonstr. 15, 12489 Berlin}

\author{     J.~Nordin}
\affiliation{    Institut fur Physik,  Humboldt-Universitat zu Berlin,
    Newtonstr. 15, 12489 Berlin}

\author{     R.~Pain}
\affiliation{ Sorbonne Universit\'e, Universit\'e Paris Diderot, CNRS/IN2P3, Laboratoire de Physique Nucl\'eaire et de Hautes \'Energies, 4 Place Jussieu, Paris, France}

\author{     E.~Pecontal}
\affiliation{   Centre de Recherche Astronomique de Lyon, Universit\'e Lyon 1,
    9 Avenue Charles Andr\'e, 69561 Saint Genis Laval Cedex, France}

\author{    R.~Pereira}
 \affiliation{    Universit\'e de Lyon, F-69622, Lyon, France ; Universit\'e de Lyon 1, Villeurbanne ;
    CNRS/IN2P3, Institut de Physique Nucl\'eaire de Lyon}

 \author{    S.~Perlmutter}
 \affiliation{    Physics Division, Lawrence Berkeley National Laboratory,
    1 Cyclotron Road, Berkeley, CA, 94720}
\affiliation{
    Department of Physics, University of California Berkeley,
    366 LeConte Hall MC 7300, Berkeley, CA, 94720-7300}

 \author{    D.~Rabinowitz}
 \affiliation{    Department of Physics, Yale University,
    New Haven, CT, 06250-8121}

 \author{    M.~Rigault}
 \affiliation{    Clermont Universit\'e, Universit\'e Blaise Pascal, CNRS/IN2P3, Laboratoire de Physique Corpusculaire,
    BP 10448, F-63000 Clermont-Ferrand, France}

 \author{    D.~Rubin}
 \affiliation{    Physics Division, Lawrence Berkeley National Laboratory,
    1 Cyclotron Road, Berkeley, CA, 94720}
    \affiliation{    Department of Physics, Florida State University,
    315 Keen Building, Tallahassee, FL 32306-4350}

 \author{    K.~Runge}
 \affiliation{    Physics Division, Lawrence Berkeley National Laboratory,
    1 Cyclotron Road, Berkeley, CA, 94720}

\author{     G.~Smadja}
\affiliation{    Universit\'e de Lyon, F-69622, Lyon, France ; Universit\'e de Lyon 1, Villeurbanne ;
    CNRS/IN2P3, Institut de Physique Nucl\'eaire de Lyon}

\author{    C.~Sofiatti}
\affiliation{    Physics Division, Lawrence Berkeley National Laboratory,
    1 Cyclotron Road, Berkeley, CA, 94720}
\affiliation{
    Department of Physics, University of California Berkeley,
    366 LeConte Hall MC 7300, Berkeley, CA, 94720-7300}

\author{    N.~Suzuki}
\affiliation{    Physics Division, Lawrence Berkeley National Laboratory,
    1 Cyclotron Road, Berkeley, CA, 94720}

\author{     C.~Tao}
\affiliation{   Tsinghua Center for Astrophysics, Tsinghua University, Beijing 100084, China }
\affiliation{    Centre de Physique des Particules de Marseille,
    Aix-Marseille Universit\'e , CNRS/IN2P3,
    163 avenue de Luminy - Case 902 - 13288 Marseille Cedex 09, France}

\author{     S.~Taubenberger}
\affiliation{    Max-Planck-Institut f\"ur Astrophysik, Karl-Schwarzschild-Str. 1,
D-85748 Garching, Germany}

\author{     R.~C.~Thomas}
\affiliation{    Computational Cosmology Center, Computational Research Division, Lawrence Berkeley National Laboratory,
    1 Cyclotron Road MS 50B-4206, Berkeley, CA, 94720}
 
\author{    M.~Vincenzi}
\affiliation{    Physics Division, Lawrence Berkeley National Laboratory,
    1 Cyclotron Road, Berkeley, CA, 94720}

\collaboration{(The Nearby Supernova Factory)}

\shortauthors{Saunders et al.}

\begin{abstract}
Type Ia supernova cosmology depends on the ability to fit and standardize observations of supernova magnitudes with an empirical model. We present here a series of new models of Type Ia Supernova spectral time series that capture a greater amount of supernova diversity than possible with the models that are currently customary. These are entitled SuperNova Empirical MOdels (\textsc{SNEMO}\footnote{https://snfactory.lbl.gov/snemo}). The models are constructed using spectrophotometric time series from $172$ individual supernovae from the Nearby Supernova Factory, comprising more than $2000$ spectra. Using the available observations, Gaussian Processes are used to predict a full spectral time series for each supernova. A matrix is constructed from the spectral time series of all the supernovae, and Expectation Maximization Factor Analysis is used to calculate the principal components of the data. K-fold cross-validation then determines the selection of model parameters and accounts for color variation in the data. Based on this process, the final models are trained on supernovae that have been dereddened using the Fitzpatrick and Massa extinction relation. Three final models are presented here: \textsc{SNEMO2}, a two-component model for comparison with current Type~Ia models; \textsc{SNEMO7}, a seven component model chosen for standardizing supernova magnitudes which results in a total dispersion of $0.100$~mag for a validation set of supernovae, of which $0.087$~mag is unexplained (a total dispersion of $0.113$~mag with unexplained dispersion of $0.097$~mag is found for the total set of training and validation supernovae); and \textsc{SNEMO15}, a comprehensive $15$ component model that maximizes the amount of spectral time series behavior captured.

\end{abstract}
\section{Introduction}

Type Ia Supernovae (SN~Ia) have provided a powerful tool for measuring the accelerating expansion of the Universe and constraining the properties of dark energy (\citealt{Riess:1998}, \citealt{Perlmutter:1999}). However, future progress will be limited by various uncertainties in their standardized magnitudes (\citealt[hearafter S15]{Saunders:2015}). While current and upcoming supernova surveys (DES, LSST, \citealt{Bernstein:2012}, \citealt{LSST:2009}) will eliminate any significant source of statistical uncertainty, along with certain calibration errors, remaining intrinsic differences in the supernovae still lead to a large amount of systematic uncertainty. Crucially, the presence of intrinsic dispersion in standardized magnitudes indicates the existence of latent unmodelled supernova processes, which can cause bias in standardized magnitudes if they are dependent on redshift or selection effects.

On the question of standardized magnitudes, progress should be possible. Type Ia supernova standardization was originally made viable by the Phillips relation (\citealt{Phillips:1993}), which identified the relation between supernova magnitudes and the decay rate of the lightcurve. This was improved further by \cite{Riess:1996} and \cite{Tripp:1998}, who added a correction to the supernova magnitude based on the difference between the $B$ and $V$ band magnitudes at maximum light, in other words the $B- V$ color. Data could be $K$-corrected with supernova templates, such as \cite{Nugent:2002fk}, and the $B$-band magnitudes could be corrected by the lightcurve width and color relations in order to be used for cosmology. Current techniques for standardization still rely on simple empirical supernova models that have only one or two degrees of freedom (\textsc{SALT2}, \citealt{Guy:2007fk}, \textsc{MLCS2k2}, \citealt{Jha:2007ys}, \textsc{SNooPy}, \citealt{Burns:2010vn}). After standardization, there is a remaining dispersion in supernova magnitudes of about $0.15$ mag (\citealt{Betoule:2014}), which is a combination of scatter due to calibration errors and intrinsic and extrinsic sources (e.g. dust) of variation in supernova magnitudes. Alternative methods for using Type Ia supernova as standard candles, such as finding ``twin" supernovae that have identical spectra (\citealt[hereafter F15]{Fakhouri:2015}) or using the near-infrared maximum magnitude (\citealt{Krisciunas:2004}, \citealt{Mandel:2011zr}, \citealt{Barone-Nugent:2012}), are restricted by the type of data required, which is not available for most existing supernova surveys.

Recent work has shown that Type Ia supernovae exhibit more diversity than can be described by two components (\citealt{Kim:2013}, S15, \citealt{Sasdelli:2015}, etc.), suggesting that the supernova-dependent sources of dispersion in the standardized magnitudes can be reduced by using an improved empirical model for Type Ia supernovae. Moreover, differences in the supernovae can be dependent on their environment, and the distribution of supernovae may shift with evolution in the environments with respect to redshift (\citealt{Kelly:2010}, \citealt{Sullivan:2010}, \citealt{Gupta:2011}, \citealt{Rigault:2013, Rigault:2015, Rigault:2018}). A model that can account for more of these differences among supernovae will reduce the potential for bias in the standardized magnitudes if the population changes with the redshift.

Since $\sim 85\%$ of a supernova's bolometric luminosity is emitted in the optical range (\citealt{Howell:2009}), a supernova model covering this range, as presented here, will cover the majority of the region where supernova variability is expressed. Further, the spectra in this optical range indicate which ions are present in the supernova ejecta, along with their velocities, as well as determining the broadband shape of the supernova lightcurves. We are motivated by the idea that these indicators uniquely determine the luminosity of the supernova, meaning that two supernovae with the same spectral time series will have the same luminosity. This hypothesis is supported by F15, where a dispersion of $0.072$~mag was attained by pairing supernovae with very similar spectral time series. (See \cite{Foley:2018} for an apparent exception) Thus, ideally, the better the supernova magnitudes can be modelled, the better they can be standardized. For these reasons, better spectrophotometric models are needed in order to extract the greatest possible amount of information from Type Ia supernovae observations.

For the empirical models currently in use, the complexity that could be incorporated in their training was constrained by the limited availability of supernova observations, in particular the small numbers of homogeneous spectral time series uncontaminated by host-galaxy light. With spectrophotometric time series data from the Nearby Supernova Factory (SNfactory, \citealt{Aldering}), a much more complex empirical model can be built. The SNfactory dataset includes hundreds of flux-calibrated, host-subtracted spectral time series which can be used to build a more complete picture of Type Ia supernovae because there is so much more information per supernova. To equivalently constrain a model with photometry, a photometric data set orders of magnitude larger and at a wide range of redshifts would be needed in order to deduce what spectral behavior drives variation in the observed broad band measurements and to break degeneracies between different spectral behavior that result in identical broad band behavior when integrated over the wavelength range of a filter. However, degeneracies between broad band colors and dust extinction, along with redshift and cosmology, may mean that a photometric data set would not be able to replace spectral data regardless of its size.

Given the nature of variations among spectra, such as the strengths and velocities of emission and absorption features, it may be that supernova spectral time series can only be perfectly described by complex nonlinear models or by explosion and radiative transfer simulations, which are currently far from obtaining the accuracy needed for supernova cosmology. However, in this analysis, we aim to make the next logical extension beyond the current generation of empirical models. The \textsc{SALT2} model is essentially performing Principal Component Analysis (PCA), finding the first two eigenvectors that describe the space of Type Ia supernova lightcurves or spectral time series. These are combined with a color relation, which is either explicitly in the form of reddening due to dust, or else is intended to describe reddening due to a combination of dust and some type of intrinsic color variation. The simplest improvement to this approach is to find the remaining significant eigenvectors, and thus build a model that describes a larger fraction of the variation in Type Ia supernovae. The data from SNfactory allows us to do this by performing PCA on a matrix made up of our collection of spectral time series. The resulting principal components, which are ranked eigenvectors of the matrix, become the linear components of a new spectral-temporal model. Using the model to fit new supernovae provides a method for standardizing their magnitudes, since correlations between the peak magnitude and fit coefficients can be calculated and corrected for. 

Other methods for modelling Type Ia supernovae have found alternative ways of reducing the dimensionality of the population space. The \textsc{SUGAR} model (L\'eget et al., in preparation), also trained on the SNfactory supernova sample, is an example of one of these methods. This model focuses on the spectral features at maximum, which strongly impact the photometric magnitudes. \textsc{SUGAR} finds the principal components in the space of spectral features, then relates those to the full spectral time series in order to build a model. Whether and in what cases this model may be preferable to the ones presented here is left to future testing.

We present the SNfactory data used here in Section~\ref{sec: Data}. The method used to calculate the model components is discussed in Section~\ref{sec: Procedure}. Three final Supernova Empirical Models (\textsc{SNEMO}) are presented in Section~\ref{sec: Final}: \textsc{SNEMO2}, a simple model to compare with \textsc{SALT2} and other current empirical supernova models, \textsc{SNEMO7}, a model chosen for standardizing supernova magnitudes, and \textsc{SNEMO15}, a model intended to capture the greatest possible amount of Type Ia supernova behavior. In this section, empirical diagnostics of the model components are measured, and the models are compared with \textsc{SALT2}. Concluding remarks are given in Section~\ref{sec: Conclusion}.

\section{The Type Ia Supernova Data Set}
\label{sec: Data}
The supernova data used in this analysis are spectrophotometric time series observed by the SNfactory between 2004 and 2014 with the SuperNova Integral Field Spectrograph (SNIFS, \citealt{Lantz:2004}). Many of these supernovae have been presented previously, for example in \cite{Thomas:2007, Thomas:2011}, \cite{Bailey:2009}, \cite{Chotard:2011}, \cite{Scalzo:2012}, \cite{Scalzo:2014}, \cite{Sasdelli:2015}, and \cite{Nordin:2018}. A publication containing the spectral time series used in this analysis is in preparation. Additionally, the supernovae in this analysis discovered by iPTF (\citealt{Kulkarni:2013}) are also presented in Papadogiannakis et al. (submitted). SNIFS is a fully integrated instrument optimized for automated observation of point sources on a structured background, i.e. the supernova host galaxy, over the full ground-based optical window at moderate spectral resolution ($R \sim600-1300$). It consists of a high-throughput wide-band pure-lenslet integral field spectrograph (IFS), an imaging channel to monitor the field in the vicinity of the IFS for atmospheric transmission variations simultaneous with spectroscopy, and for acquisition and guiding. The IFS possesses a fully-filled $6.\!''4 \times 6.\!''4$ spectroscopic field of view subdivided into a grid of $15 \times 15$ spatial elements, a dual-channel spectrograph covering $3200-5200$~\AA\ and $5100-10000$~\AA\ simultaneously, and an internal calibration unit (continuum and arc lamps). SNIFS is continuously mounted on the South bent Cassegrain port of the University of Hawaii $2.2$ m telescope on Mauna Kea and is operated remotely. A description of host-galaxy subtraction is given in \cite{Bongard:2011}. 

The supernovae are flux-calibrated following the procedure described in \citet{Buton:2012}, using a system of both CALSPEC (\citealt{Bohlin:2014} and references within) and Hamuy (\citealt{Hamuy:1992, Hamuy:1994}) standard stars, which are calibrated on the \cite{Hayes:1975} system and listed in \citet{Buton:2012}. Calibration of the data on non-photometric nights is described in \cite{Pereira:2013}. Note that this hybrid calibration system is mildly in tension with data calibrated purely with the CALSPEC standards (an effect seen earlier in S15). This calibration difference primarily affects the $U$-band; the effect on the $B-V$ color is minimal but the $U-B$ color can be approximately $0.02$ mag lower in the system used here. The redshifts of the supernovae are given in \cite{Childress:2013} and Rigault et al. (\citealt{Rigault:2018}). The supernovae have been corrected for Milky Way dust extinction (\citealt{Schlegel:1998}, \citealt{Cardelli:1989}). The wavelengths have been shifted to a common restframe at $z=0$. All references hereafter to the phase or wavelength of the supernova data will refer to the values in this restframe.

The supernovae used in this analysis range in redshift from $z=0.01$ to $0.08$. They are required to have observations on at least $5$ epochs, with at least one epoch between $10$ days before and $7$ days after maximum light, as determined by fitting the data with \textsc{SALT2}, and $4$ epochs between $10$ days before and $35$ days after maximum light. Stricter phase-based cuts on the data set are also discussed below. In this analysis, we include peculiar supernovae, such as SN1991T and SN1991bg-like objects, plus other supernovae that are not well fit by the \textsc{SALT2} model. Whether and how to exclude such supernovae from a cosmology analysis are discussed in Section~\ref{sec: mag std} and in the Appendix. The SNfactory data set used here has $223$ supernovae that meet these data cuts. Future versions of the SNfactory data production are expected to increase this number.

The date of maximum fit by \textsc{SALT2} is used to align the supernovae by phase in the analysis that follows. This phase is well constrained because of the other constraints put on the data set, and its uncertainty is generally well under half a day even for supernovae that are not very well matched by the \textsc{SALT2} model. This small level of uncertainty is confirmed in the final models, which give a small scatter ($\textsc{RMS}=0.3$ days) in the date of maximum of the reconstructed supernovae. Additionally, in the final data set used in this analysis only supernovae with data before maximum are used, further constraining the date of maximum.

\section{Method for Building the Supernova Models}
\label{sec: Procedure}
The goal of the analysis is to build a model for the time-evolving spectral energy distribution of Type~Ia supernovae that will consist of spectral time series components. There are a number of reasons for using a model for Type Ia supernovae, and we present separate analysis products to address these: a simple model with two components (the first of which is essentially the average spectral time series) plus a color relation to compare our analysis method with \textsc{SALT2}; a model optimized to work best for standardizing supernova absolute magnitudes; and a model intended to best match the spectral time series diversity of all Type Ia supernovae.

Following the precedent of current empirical supernovae models, each of the produced models will consist of a number of spectral time series components. For each model, the components are linearly combined to match the spectral time series of a given supernova. The spectral-temporal model components are produced by first regressing the supernova data onto a regular grid in phase and wavelength, discussed in Section~\ref{sec: gp}, then calculating the principal components of this regularly sampled data, discussed in Section~\ref{sec: emfa}.

In training these empirical models, a number of choices must be made. First, we consider the selection criteria for the model training supernova set, namely whether supernova used in training the model must have epochs observed at early or late times. Also considered is whether the supernova spectra should be dereddened to account for dust between the supernova and the observer as a separate step and what dust reddening relation to use, discussed in Section~\ref{sec: color}. Lastly, the number of model components must be determined for the model optimized for magnitude standardization and the comprehensive spectral time series model. To avoid overfitting, these decisions are made using K-fold cross-validation, described in Section~\ref{sec: Model Testing}.

\subsection{Regression of Supernova Observations onto a Regularly Sampled Grid in Wavelength and Phase}
\label{sec: gp}
The supernova data from the SNfactory are by necessity not uniformly sampled in phase and have varying degrees of noise due to the redshift range, phase, and observing conditions of the supernovae. To facilitate the calculation of principal components, Gaussian Processes are used to regress the irregular data onto a uniform grid in wavelength and phase. Gaussian Processes (Rasmussen and Williams, 2006) assumes that an observation is a sample drawn from some underlying function with an error model: 
\begin{equation}
y = \mathbf{f}(x) + \epsilon 
\end{equation}
where $\epsilon$ is a normal distribution that accounts for the error in the data. A Gaussian Process models the underlying function as
\begin{equation}
\mathbf{f} \sim \mathcal{N}(a \;  m(X), K(X,X))
\end{equation}
where $X$ are the input dependent variables, here two-dimensional elements corresponding to the wavelength and phase for each flux measurement, $m(X_{\star})$ is the model for the mean function, $a$ is a fit normalization parameter, and $K$ is a covariance matrix made up of Mat\'ern kernel (\citealt{Matern:1986}) elements 
\begin{equation}
k(x, x') = \sigma_{f}^{2} \Bigg[ 1 + \frac{\sqrt{5}(x-x')}{l} + \frac{5 (x-x')^{2}}{3l^{2}} \Bigg] \exp{ \Bigg( -\frac{\sqrt{5}(x - x')}{l} \Bigg)} + \sigma_{n}^{2}\delta(x,x')
\end{equation}
where $l$, $\sigma_{f}$, and $\sigma_{n}$ describe the correlation scale between variables, the amount of variation allowed from the mean, and the independent variation in the data.

In this analysis, the mean function $m(X)$ is constructed by taking the smoothed and normalized mean of all the supernovae, with the \cite{Hsiao:2007uq} template as a prior at epochs where there is little data. The correlation scale $l$ is fit globally for all supernovae and is only applied to data points with different phases but the same wavelength. No correlation is enforced between elements at different wavelengths since the data has already been binned in wavelength and enforcing a correlation was not found to improve results. The other parameters ($a$, $\sigma_{f}$, and $\sigma_{n}$) are fit for each supernova by maximizing the log-likelihood of the model. These parameters are fit for each individual supernova because their values relate to the observing conditions and the amount the supernova differs from the average.

Given input data at $X$ with variance $V$, the prediction for the underlying supernova spectrum at points $X^{\prime}$ is then the mean of the conditional distribution of function values drawn from the normal distribution:
\begin{equation}
\label{eq: GP}
\mathbf{f^{\prime}} = a\; m(X^{\prime}) + K(X^{\prime}, X)[K(X,X)+V]^{-1}(\mathbf{y} - a \;m(X))
\end{equation}
with covariance
\begin{equation}
\label{eq: GPcov}
cov(\mathbf{f^{\prime}}) = \mathbf{\Sigma^{\prime}} = K(X^{\prime}, X^{\prime}) - K(X^{\prime}, X) \ [K(X,X) + V]^{-1} \ K(X,X^{\prime}).
\end{equation}

\begin{figure}
\includegraphics[width=\columnwidth]{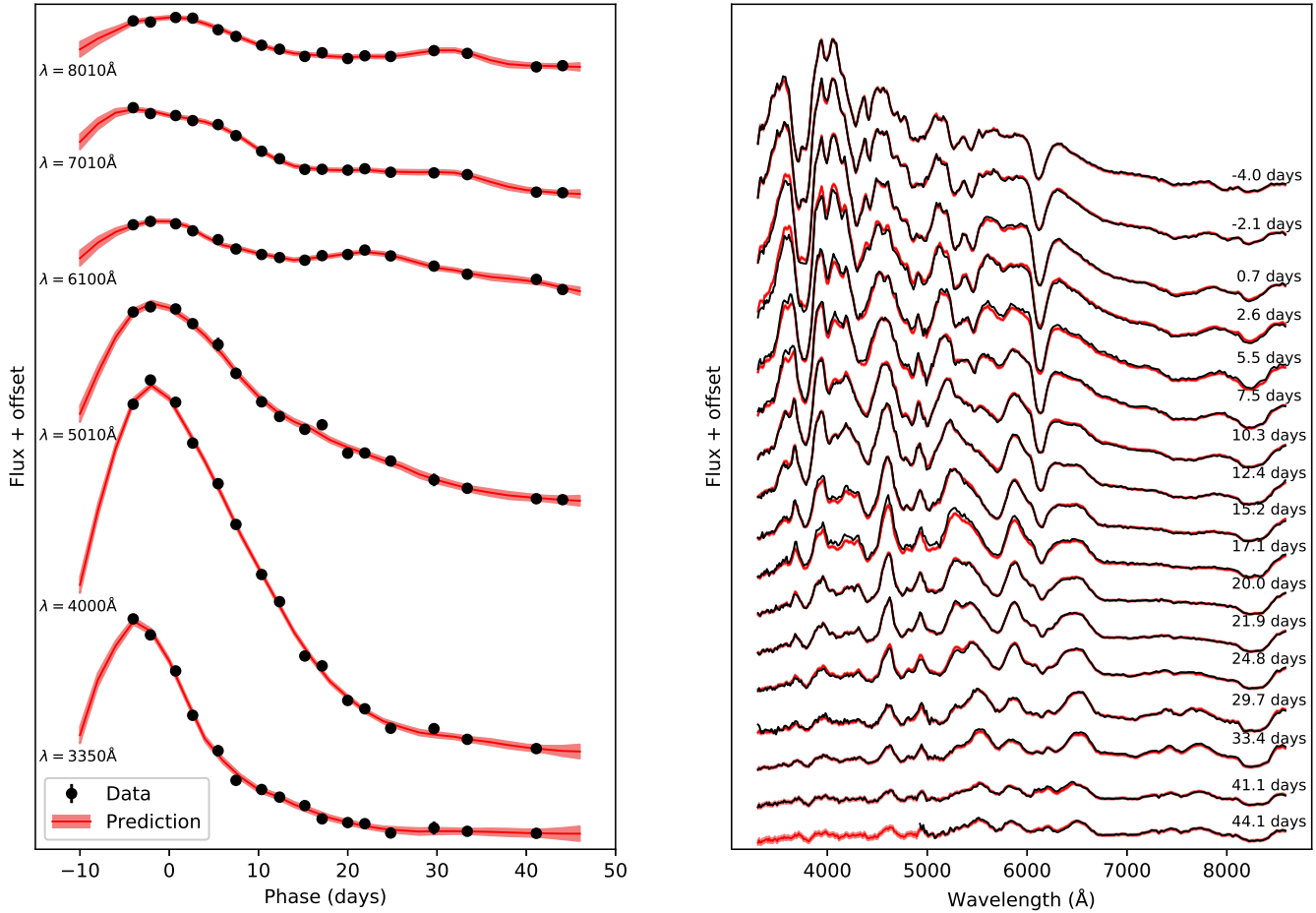}
\caption{A demonstration of the supernova model predicted by Gaussian Processes compared with the supernova data. The left panel shows monochromatic lightcurves while the right panel shows the spectra predicted by the Gaussian Processes at the phases where there is supernova data.}
\label{fig: GP lc demo}
\end{figure}

\begin{figure}
\includegraphics[width=0.9 \columnwidth]{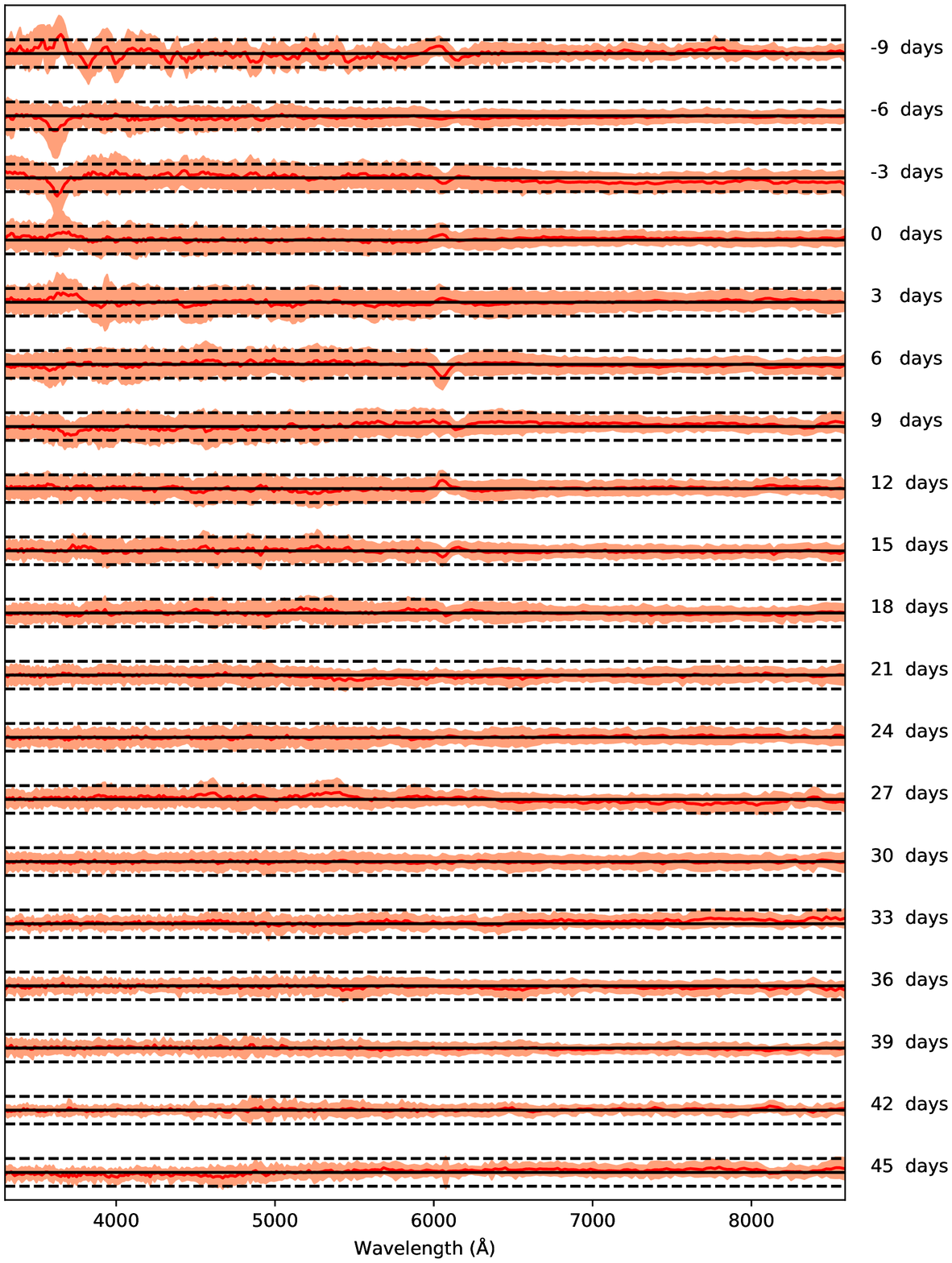}
\centering
\caption{Average and standard deviation of the pulls between Gaussian Process predictions and the data versus the wavelength, in $3$ day phase bins. The red lines show the average pull, which is defined as the residual divided by the error, of the Gaussian Process prediction to the data at the phase indicated to the right of the figure, while the shaded region shows the standard deviation in the pulls. The solid black lines are the zeropoint for the averages, where a level of zero indicates that there is no bias. The dashed lines show the level of plus or minus one in the pulls.}
\label{fig: GP bias}
\end{figure}

For each supernova in our data set, a prediction is made in this manner for the true spectral time series in flux of the object. For this analysis, we generate spectral time series spanning $10$ days before maximum and $46$ days after maximum at two day intervals, with $288$ wavelength elements between $3300$--$8600 \mathrm{\AA}$ at a constant velocity resolution of $1000$ km/s. These phase and wavelength ranges are determined by the range in which the SNfactory supernovae are well observed, which correspond to the phase and wavelength range in which the majority of supernova luminosity is emitted (\citealt{Howell:2009}). Fluxes here and after are $f_\lambda$ ($\mathrm{ergs} \; \mathrm{s}^{-1}\; \mathrm{cm}^{-2} \; \mathrm{\AA}^{-1}$). Figure~\ref{fig: GP lc demo} shows an example of the Gaussian Process prediction for a moderately well-sampled supernova with some scatter in the data. Figure~\ref{fig: GP bias} shows the performance of the predictions over the whole supernova set, demonstrating that the predictions are unbiased and have residuals consistent with the estimated uncertainty on the data around maximum light. The uncertainty at later times is somewhat overestimated, which can be attributed to the fact that the matrix $K$ is fixed for all epochs of the supernova and over-estimates the variability at later times. A more advanced epoch-dependent model might be able to account for this, but this is left for future work.

\subsection{Dereddening the spectra to account for interstitial dust}
\label{sec: color}
Supernova color differences have been shown to be a major contributor to inhomogeneity in supernova spectra and magnitudes (\citealt{Riess:1996}, \citealt{Tripp:1998}). This color is some combination of intrinsic color and reddening due to dust, but a generally accepted method for separating these components has not yet been discovered (\citealt{Mandel:2014}, Kim et al., submitted). The portion due to dust reddening is expected to obey an extinction relation, such as in \citet[hereafter C89]{Cardelli:1989}, or another similar model such as \citet[hereafter FM07]{Fitzpatrick:2007}. Dust reddening is well described by an exponential multiplier to the flux and thus cannot be described by a single linear principal component. Dereddening the supernovae using the C89 relation before calculating the spectral time series model components may lead to a model that better matches supernova behavior than a model without dereddening and means that a smaller basis of linear components should be needed to describe supernova variability. Thus, in addition to calculating a purely linear supernova model, we also calculate models using supernova data that are first dereddened with the C89 or the FM07 dust reddening models. Supernova dereddening was seen to be dependent on the phases of the spectra used, for example when only early phases, which are seen to be highly variable in time, or only more uniform later phases, are available to calculate the dereddening. Because of this, the dereddening process is done using the Gaussian Process prediction for the supernova, instead of the raw spectra, so that all phases are included. 

The color relation is applied by fitting the color difference between each supernova and a fiducial supernova and dereddening the supernova by that amount. This is done by minimizing the quantity 
\begin{equation}
\label{eq: dered}
\sum_{i=1}^{N_{SNe}}
\Big(\mathbf{f_{SN_i}}(p, \lambda) - a_{i} \; 10^{-0.4\; A_{s, i} \; \mathbf{C}(\lambda)} \; \mathbf{f_{fid}}(p, \lambda)\Big)^{\textrm{T}}\mathbf{\Sigma_{SN_i}}^{-1}\Big(\mathbf{f_{SN_i}}(p, \lambda) - a_{i} \; 10^{-0.4\; A_{s, i} \; \mathbf{C}(\lambda)} \; \mathbf{f_{fid}}(p, \lambda)\Big)
\end{equation}
as a function of $a_{i}$, $A_{s, i}$, and $\mathbf{f_{fid}}(p, \lambda)$, where $A_{s, i}$ is the extinction difference between $SN_{i}$ and the fiducial supernova, $a$ is a normalization factor, $\mathbf{f_{fid}}(p, \lambda)$ is a freely fit spectral time series varying as a function of the phase $p$ and wavelength $\lambda$, and $\mathbf{f_{SN_i}}$ and $\mathbf{\Sigma_{SN_i}}$ are the predicted flux and covariance for each supernova from the Gaussian Processes step (Equations~\ref{eq: GP} and \ref{eq: GPcov}). $\mathbf{C}(\lambda)$ is either the C89 or FM07 dereddening model, both using $R_{V} = 3.1$. (The degeneracy between $a_{i}$, $A_{s, i}$, and $\mathbf{f_{fid}}(p, \lambda)$ is removed by fixing the value for $a_{i}$ for an arbitrarily chosen supernova and fixing the average of $A_s,i$ to zero.) A fit fiducial supernova is used, rather than a template such as \citet{Hsiao:2007uq}, in order to ensure that the maximum amount of variation due to dust-like reddening allowed by the data is removed.

The supernovae dereddened using the fit $A_{s}$ values (but not including the normalization factor $a$) are then used as input to the calculation of the components that will form the final spectral-temporal model. For brevity, $\mathbf{F_{SN}}(p, \lambda)$ will refer hereafter to the dereddened spectral time series of the supernova, i.e. 
\begin{equation}
\mathbf{F_{SN}}(p, \lambda) = \mathbf{f_{SN}}(p, \lambda) 10^{0.4 \; A_{s} \; \mathbf{C}(\lambda)}
\end{equation}

\subsection{Extracting Principal Components from the Spectral Time Series}
\label{sec: emfa}
In order to calculate linear components from the supernova spectral time series predictions produced using Gaussian Processes, we use a variation of Principal Component Analysis (PCA) that is capable of incorporating data with uncertainties. This variation, Expectation Maximization Factor Analysis (EMFA) is a method for performing dimensionality reduction presented in \citet{Ghahramani:1996}. In addition to utilizing data uncertainty, EMFA has the added advantage of allowing the user to choose the number of model components that will be used to describe the data set, in other words the basis of $N$ vectors that best fits a given space. Note that this means that a model with $N$ components is not a subset of a model with $N+1$ components trained on the same data set.

EMFA models the given data as 
\begin{equation}
\mathbf{F} = \mathbf{e} \; \mathbf{c} + \mathbf{u}
\end{equation}
where $\bold{F}$ is the $N_{variables} \times N_{SNe}$ dimensional data matrix (here the Gaussian Process predictions for each phase and wavelength combination) and $\mathbf{e}$ is a $N_{variables} \times N_{vectors}$ dimensional matrix of model components. The factors $\bold{c}$ (forming a $N_{vectors} \times N_{SNe}$ dimensional matrix) correspond to the coefficients of the model. $N_{vectors}$ is generally much smaller than $N_{variables}$. Here we have $8352$ variables corresponding to $288$ wavelength elements times $29$ phase elements, while we predict something on the order of $10$ vectors. Finally, $\bold{u}$ is a $N_{variables} \times N_{SNe}$ dimensional random variable with distribution $\mathcal{N}(0, \mathbf{\Sigma})$, corresponding to the noise $\mathbf{\Sigma}$ in $\bold{F}$, and is here taken from the covariance predicted in the Gaussian Processes step (see Section~\ref{sec: gp}). While the method is geared towards data centered around zero, here we do not subtract the mean so as to prevent introducing assumptions about the absolute flux of the supernovae (in terms of calibration and cosmology) into the model. Testing has shown that the method is stable without subtracting the mean and is also insensitive to moderate changes in the normalization of the supernovae, though it starts to degrade if order-of-magnitude variations are introduced in the brightness scale of the data. The method of \citet{Ghahramani:1996} has been adapted here to accommodate correlated uncertainties.

The method consists of an iterative process in which the expected value of the factor $\bold{c}$ is calculated, given $\mathbf{e}$, and then new values for $\mathbf{e}$ are calculated given the data and the new values for $\bold{c}$. This is repeated until the log likelihood converges, here determined by the point where the percentage change in the log likelihood is less than $10^{-7}$. Once convergence is reached, the components in the matrix $\mathbf{e}$ can be orthogonalized and ranked using a standard Singular Value Decomposition algorithm. 

The resulting model for a supernova takes the form
\begin{equation}
\mathbf{F}_{SN} (p, \lambda) = \sum_{i=0}^{N-1} c_{SN, i} \ e_{i}  (p, \lambda)
\end{equation}
where $\mathbf{F}_{SN}$ is the flux of the supernova at the phase $p$ and wavelength $\lambda$, $N$ is the number of model components, $e_{i}$ are the model components (the eigenvectors), and $c_{SN, i}$ are the model coefficients fit to the individual supernova. Since the model is expected to be used for supernovae at a range of redshifts, we rewrite the model in the form
\begin{equation}
\label{eq: sn model}
\mathbf{F}_{SN} (p, \lambda) = c_{SN, 0}\Big(e_{0}(p, \lambda) + \sum_{i=1}^{N} c'_{SN, i} e_{i}  (p, \lambda)\Big)
\end{equation}
where $c'_{SN, i} = \frac{c_{SN, i}}{ c_{SN, 0}}$, so that $c_{SN, 0}$ is the only model coefficient that is dependent on the flux normalization and is thus the only coefficient that is a function of the luminosity distance. Because of the leading multiplicative coefficient, the model is not affected by the flux normalization of the training data. Hereafter, the prime is dropped and $c'_{SN, i}$ is referred to as simply $c_{SN, i}$. Note that this step creates correlations between the model coefficients, even though the eigenvectors are orthogonal.

\subsection{Making Decisions on the Model Training Using Cross-Validation}
\label{sec: Model Testing}
As described above, there are a number of decisions to make regarding the training of the models. The first question is whether to make any cuts on the supernova set (in addition to those described in Section~\ref{sec: Data}) based on the phase coverage of the observations. For this, we compare the model components calculated using the full data set to a set only including supernovae that have observations before maximum, which brings the total number of supernovae to $171$, and a set only including supernovae that have observations before maximum and at least $30$ days after maximum, comprising $140$ supernovae. The number of supernovae in each level of cuts is given in Table~\ref{table: sets}. Second, as described in Section~\ref{sec: color}, we consider whether to deredden the supernovae prior to calculating the model components and whether to use the reddening relation of FM07 or C89. Last, we must determine how many components to use in the model optimized for standardizing supernova magnitudes and the model optimized to fully capture the spectral time series for all Type Ia supernovae.

\begin{deluxetable}{l c c c c }
\label{table: sets}
\tablecolumns{5}
\tablecaption{Numbers of Supernovae in each of the Levels of Data Selection}
\tablehead{ \colhead{Data Selection} & \colhead{Each Training Set} & \colhead{Each Test Set} & \colhead{Final Hold-out Set} & \colhead{Total}}
\startdata
All Supernovae & 120 & 39 & 64 & 223 \\
Supernovae with pre-max data & 92 & 30 & 49 & 171 \\
Supernovae with pre-max and post-30 days data & 77 & 25 & 38 & 140 \\
\enddata
\end{deluxetable}

Cross-validation is used to determine the best option for each of these issues. After withholding $\sim30\%$ of the total number of supernovae randomly selected to be used as a final demonstrative set, the remaining supernovae are divided into $4$ training and test set pairs, following a k-fold cross validation procedure where the model is trained on $3/4$ of the data and tested on the last $1/4$. The options that best match the supernovae are chosen by looking at the average of given metrics (described below) over the $4$ sets. Once the options have been chosen in this step, the final models are trained accordingly on all of the supernovae (excepting the supernovae withheld as a final demonstration set).

\subsubsection{Selecting the phase coverage criterion and dereddening procedure}
To select the training set phase coverage criterion and method of accounting for reddening, we first consider how well the models are able to fit the test set supernovae. To do this, we calculate the value
\begin{equation}
\chi^2 (n) = \sum^{SNe}_{i=0} \Big( \mathbf{F^{n}_{i}} (p, \lambda) - \mathbf{F_{SN_i}}(p, \lambda) \Big)^{\textrm{T}} \mathbf{\Sigma_{SN_i}}^{-1}  \Big( \mathbf{F^{n}_{i}} (p, \lambda) - \mathbf{F_{SN_i}}(p, \lambda) \Big)
\end{equation}
where $\mathbf{F}^{n}_{i}$ is the $n$-component model fit to supernova $i$, and $\mathbf{F_{SN_i}}$ and $\mathbf{\Sigma_{SN_i}}$ are the predicted flux and covariance from the Gaussian Processes step. The average of the reduced $\chi^2$, in other words the $\chi^2$ per degree of freedom, over each of the $4$ test sets is compared for each of the combinations of training sets and dereddening options. Reduced $\chi^2$ values below $1$ are driven by the overestimated uncertainties at later epochs from the Gaussian Processes step, discussed in Section~\ref{sec: gp}. Since we are concerned with the relative value of the $\chi^2$ for different models using the same data, the absolute value of $\chi^2$ is not a concern here. We find that the models using the FM07 reddening relation and trained with supernovae that have pre-maximum observations are equivalent or slightly preferred to the others, while also performing best on the metrics considered in the following sections. As preference for the Fitzpatrick reddening relation was also seen in recent studies, such as \cite{Schlafly:2010} and \cite{Huang:2017}, we select this dereddening procedure for the models.

\begin{figure}
\includegraphics[width = \columnwidth]{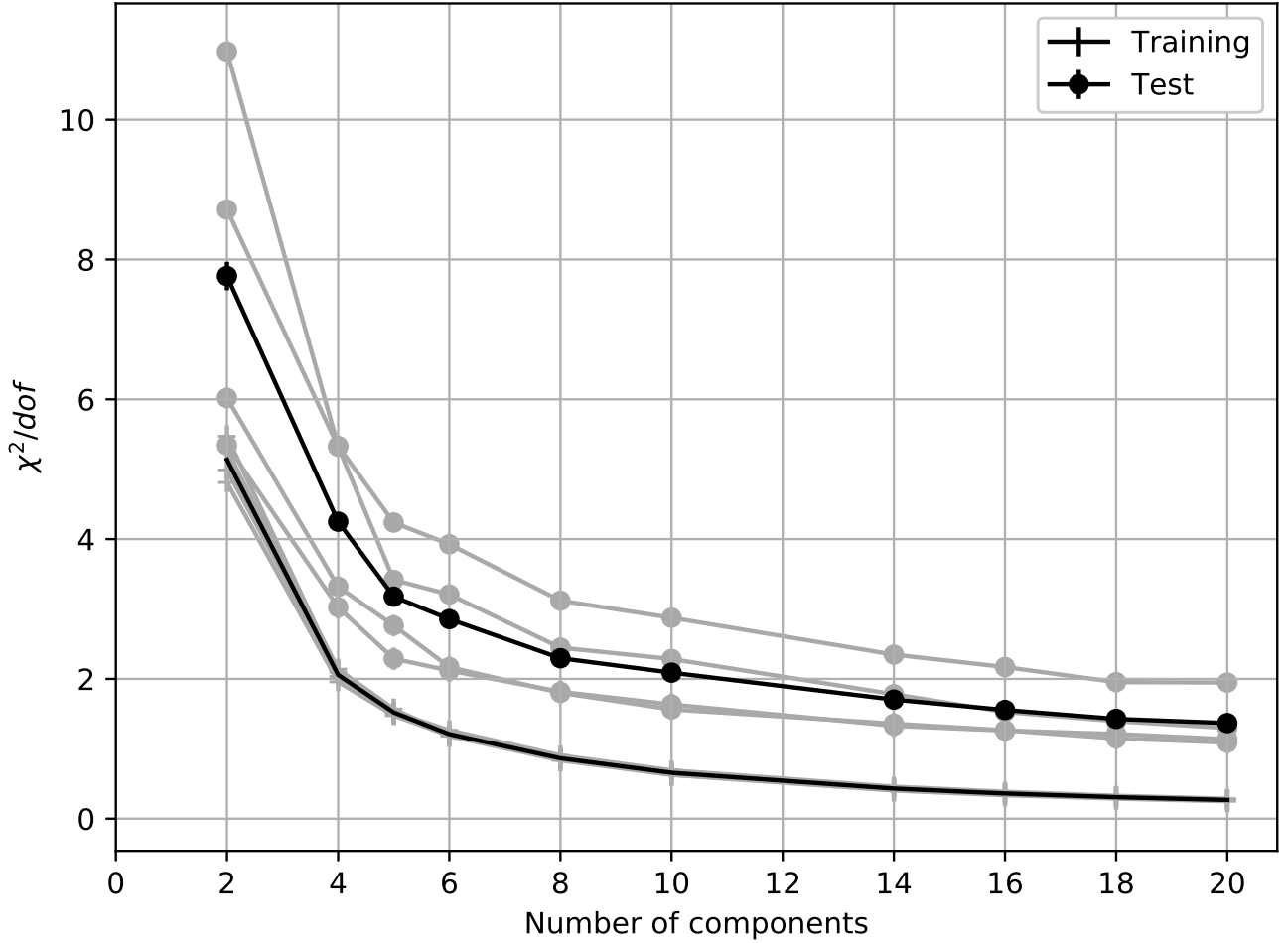}
\caption{Reduced $\chi^2$ values for supernovae with prior dereddening using the FM07 extinction curve versus the number of model components. Gray lines show the results for each of the cross-validation training and test sets, while black lines show the median and error on the median for the training and test sets.}
\label{fig: Chi2-FM07}
\end{figure}

The $\chi^2$ metric, shown in Figure~\ref{fig: Chi2-FM07}, does not provide a way to choose the optimal number of components to use for the model since the reduced $\chi^2$ value follows roughly asymptotic behavior for the test sets and does not have a minimum indicating the number of components preferred by the data.

\subsubsection{Selecting the number of components for the magnitude standardization model}
\label{sec: mag std}
In terms of its utility for cosmology, the primary metric for a supernova model is how well it can standardize supernova magnitudes and minimize their unexplained dispersion, that is, the combination of effects leading to dispersion in the supernova magnitudes that are not captured by our model. To select the model that leads to the lowest unexplained dispersion, we perform a magnitude standardization procedure on each model following the method of the `Union' supernova compilation (\citealt{Amanullah:2010}). The model chosen is the one for which the unexplained dispersion is lowest. This procedure performs a global fit for the level of the unexplained dispersion simultaneously with the parameters $\boldsymbol{\alpha}$, $\alpha_{c}$ and $M_{g}$ used to calculate each supernova's distance modulus:
\begin{equation}
\mu_{g} = m_{g}^{std} - M_{g}= m_{g} + \sum_{i} \alpha_{i} \; c_{i} \; +\;\alpha_{c} \times A_{S} - M_{g}
\end{equation}
where $c_i$ and $A_S$ are the model parameters, determined by the previously-obtained fits for the supernovae from Equation~\ref{eq: sn model}, $m_{g}$ is SDSS $g$-band magnitude at maximum obtained by performing synthetic photometry on the model formed by those fit parameters, $M_{g}$ is the standardized absolute magnitude of the supernova and $\mu$ is the distance modulus. In the case where no prior dereddening step is performed, the $\alpha_{c} \times A_{S}$ term is omitted. In this procedure we also reject supernovae that are outliers in the model coefficient space, as described in the Appendix, because in this section only a linear standardization is assumed. The supernovae identified in this way were already known to be peculiar (\citealt{Scalzo:2012}), though this knowledge was not used to select them as outliers in this analysis.

The standardization parameters are fit using an iterative procedure in which the log-likelihood function
\begin{equation}
\mathcal{L} =  \sum_{i}^{N_{SNe}} \bigg\{-\ln[2\pi \;(\sigma^{2}_{pv, i}+ \sigma^{2}_{m, i} + \sigma^2_{u})]  -\frac{[ \mu_{g}(\boldsymbol{\alpha}, \alpha_{c}, M_{g}) - \mu(z_{i}, \theta) ]^{2}}{\sigma^{2}_{pv, i} + \sigma^{2}_{m, i} + \sigma^2_{u}} \bigg\}
\label{eq: disp}
\end{equation}
is maximized with a fixed value of the unexplained dispersion $\sigma_{u}$, and then the value of the unexplained dispersion is fit such that the reduced $\chi^2$ is equal to unity. The distance modulus $\mu(z, \theta) \approx 5 \log(1+z) + constant$, where $\theta$ represents the parameters of the cosmological model, is insensitive to the cosmological model since the supernova redshifts are very low and the constant is degenerate with the supernova absolute magnitude $M_g$; $\sigma_{pv}$ accounts for uncertainty due to host galaxy peculiar velocities of $300 \;\textrm{km s}^{-1}$; and $\sigma_{m}$ is the uncertainty in the fit of the supernova model,
\begin{equation}
\sigma_{m} = \mathbf{V}^{T} \mathbf{C} \mathbf{V} \qquad\textrm{where}\qquad \mathbf{V} = \left ( \begin{array}{c} 1 \\ \alpha_c \\ \boldsymbol{\alpha} \end{array} \right )
\end{equation}
and $\mathbf{C}$ is the covariance in the model fits to the supernova data.

It is important to note that this standardization method is purely empirical and not an analytic consequence of the model. It is based on finding empirical correlations between the $c_{i>0}$ components, which are independent of the luminosity distance (see Equation~\ref{eq: sn model}), and the magnitudes of the supernovae in the data set used here, thus separating the intrinsic and distance-dependent parts of the supernova magnitude. More advanced techniques, potentially requiring more data, could be used to improve upon the linear standardization performed here (discussed further in Section~\ref{sec: disp final}).

\begin{figure}
\includegraphics[width =\columnwidth]{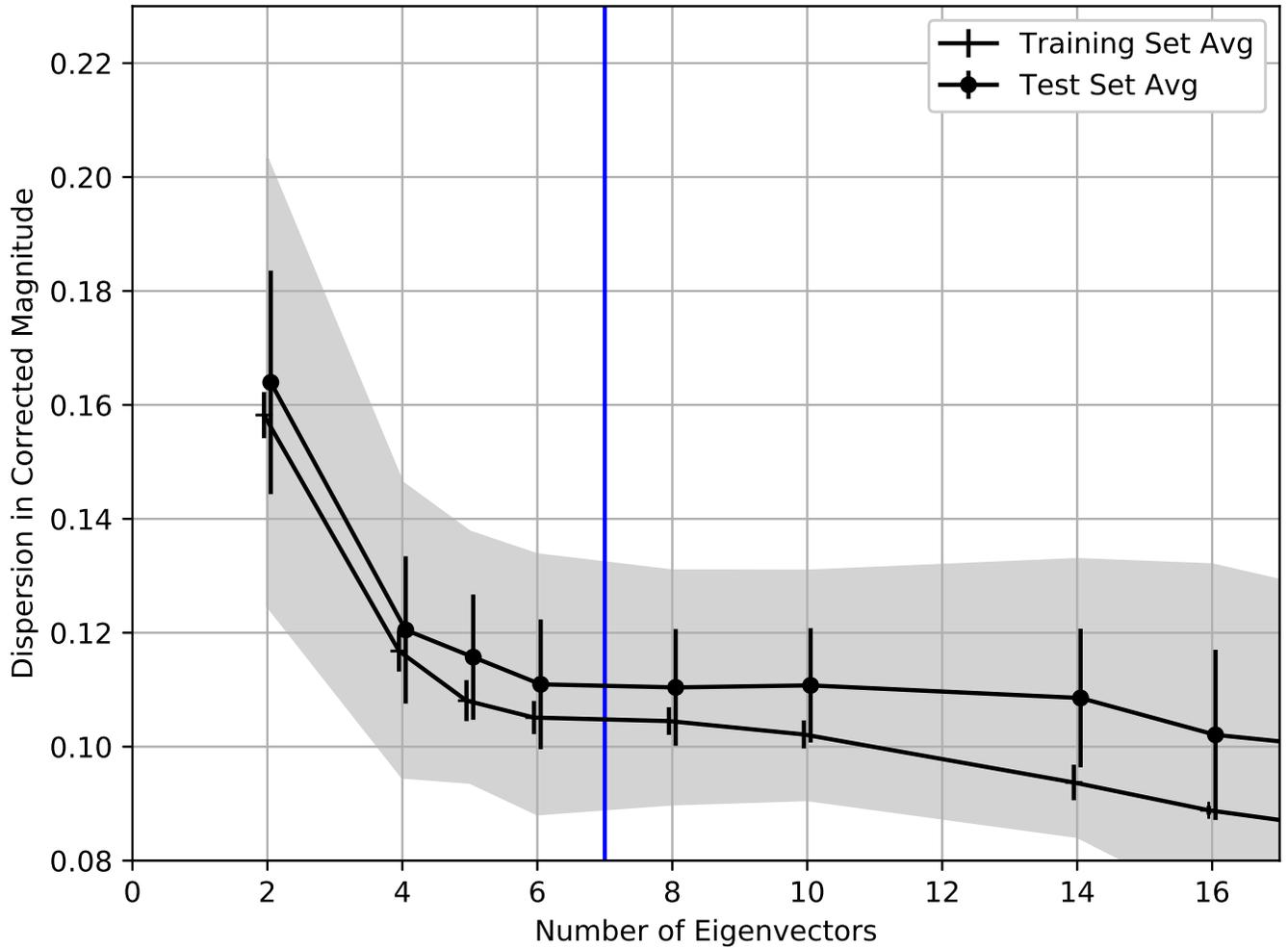}
\caption{Unexplained dispersion in standardized magnitudes for supernovae with FM07 dereddening versus the number of model components. The solid lines show the result averaged over all of the $K$-fold sets and the error bars shows the error on the mean, while the shaded region shows the standard deviation among the test sets. Points are offset slightly on the $x$-axis to facilitate interpretation. The vertical blue line shows the local minimum at $7$ vectors.}
\label{fig: Disp Results-FM07}
\end{figure}

The standardization parameters are calculated from the training set, then applied to the test set. To select the number of components to use for the model optimized to minimize dispersion in supernova magnitudes, we choose the point where the unexplained dispersion stops improving for the test set, shown in Figure~\ref{fig: Disp Results-FM07}. Fitting a curve to the modelled points gives a local minimum at $7$ components. While the unexplained dispersion decreases again at $16$ components, we choose the minimum at $7$ components as this is more likely to be constrainable with photometric data, which is the target use case for this model.

\subsubsection{Selecting the number of components for the comprehensive supernova model}
\begin{figure}
\includegraphics[width = \columnwidth]{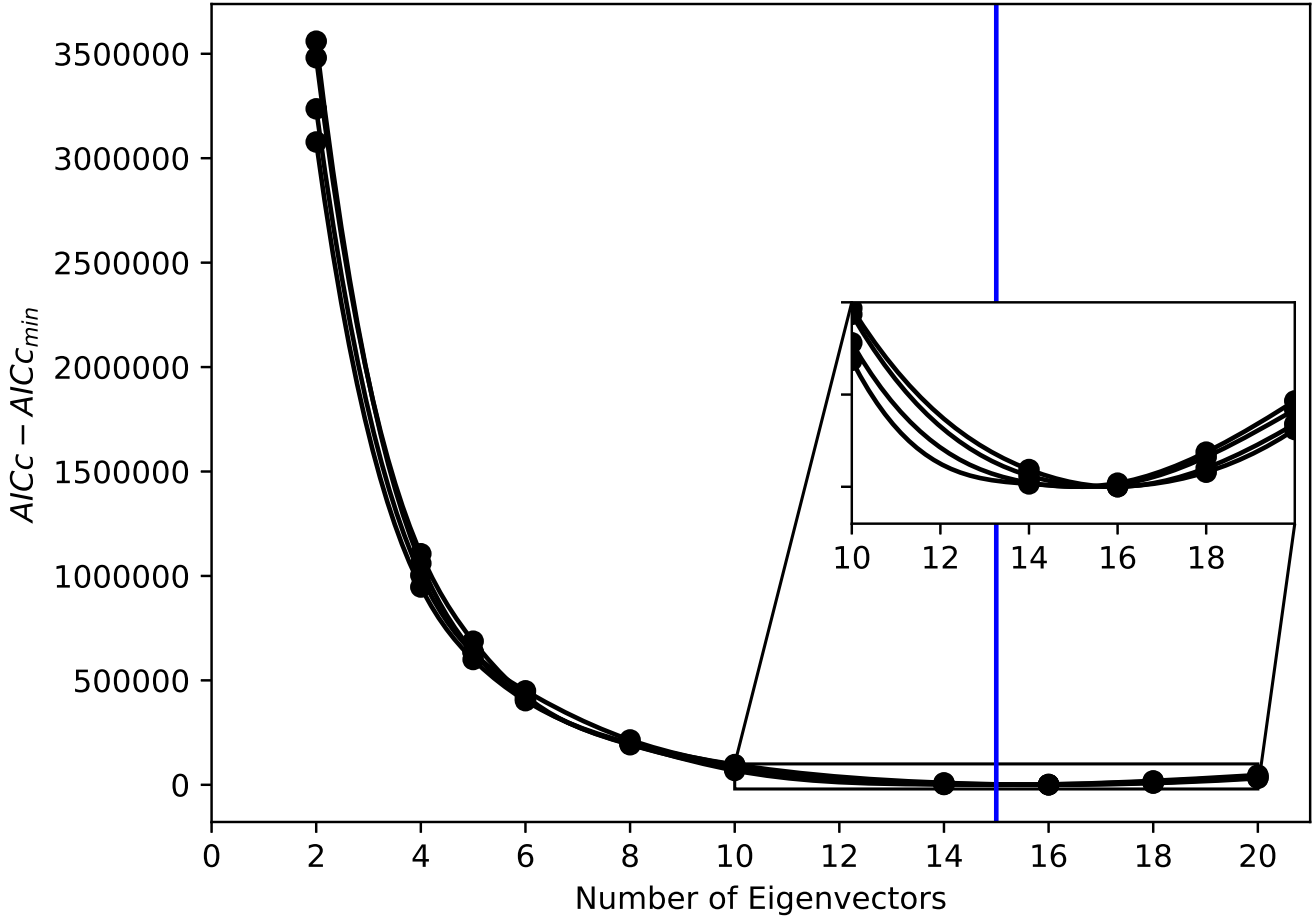}
\caption{The AIC metric applied to the models trained on the supernovae with data before maximum, corrected with the FM07 reddening relation, versus the number of model components. The four lines correspond to the different cross-validation training sets. The vertical blue line shows the minimum at $15$ vectors.}  
\label{fig: AIC}
\end{figure}

To select the model with the number of components that best describes the full set of supernovae we compare the Akaike Information Criterion (AIC, \citealt{burnham2003}) for each number of model components. This method measures the relative quality of models for a given set of data and puts a stronger penalty on adding complexity to the model than the reduced $\chi^2$ metric. Because of the fact that a very large number of variables (corresponding to the number of wavelength and phase elements of the spectral time series) are added for each additional model component, the models are not well suited to calculating the relative probability between different numbers of components, but a minimum AIC value can be found. The AIC metric versus the number of model components is shown in Figure~\ref{fig: AIC}, where the minimum at $15$ components indicates that this model best describes the supernova population.

\section{The Final Spectral-Temporal Models}
\label{sec: Final}
With all training decisions made as described in the previous section, the $2$, $7$, and $15$ component models, called \textsc{SNEMO2}, \textsc{SNEMO7}, \textsc{SNEMO15}, are calculated with the full training set of $122$ supernovae with observations before maximum (still reserving $49$ for use in the hold-out final demonstration set). The resulting model components are given in Table~\ref{tab:alleigenvs} and the corresponding coefficients for the training and hold-out sets of supernovae are given in Table~\ref{tab:allcoeffs}.
\begin{deluxetable}{ccccccc}
\tablewidth{0pc}
\tablecolumns{7}
\tablecaption{Components for the SNEMO Models\label{tab:alleigenvs}}
\tablehead{
\colhead{Phase (days)} & \colhead{Wavelength ($\mathrm{\AA}$)} & \colhead{SNEMO2 $\mathbf{e}_{0}$} & \colhead{SNEMO2 $\mathbf{e}_{1}$} & \colhead{SNEMO7 $\mathbf{e}_{0}$} & \colhead{SNEMO7 $\mathbf{e}_{1}$} & \colhead{SNEMO7 $\mathbf{e}_{2}$}  }
\startdata
-10.00 & 3305 & 1.201 & -0.2772 & 1.365 & 0.7203 & 0.03686\\
-10.00 & 3317 & 1.235 & -0.2931 & 1.404 & 0.7471 & -0.01845\\
-10.00 & 3328 & 1.271 & -0.2961 & 1.448 & 0.7512 & -0.04672\\
-10.00 & 3339 & 1.313 & -0.2654 & 1.485 & 0.7356 & -0.09023\\
-10.00 & 3350 & 1.306 & -0.2396 & 1.482 & 0.7120 & -0.05760\\
-10.00 & 3361 & 1.294 & -0.2407 & 1.489 & 0.7122 & -0.04199\\
\enddata
\end{deluxetable}

\begin{deluxetable}{lccccc}
\tabletypesize{\footnotesize}
\tablewidth{0pc}
\tablecolumns{6}
\tablecaption{Training and Validation Set Model Parameters\label{tab:allcoeffs}}
\tablehead{
\colhead{SN Name} & \colhead{SNEMO2 $c_{0} - \bar{c_{0}}$} & \colhead{SNEMO2 $c_{1}$} & \colhead{SNEMO7 $c_{0}- \bar{c_{0}}$} & \colhead{SNEMO7 $c_{1}$} & \colhead{SNEMO7 $c_{2}$}  }
\startdata
Train\textunderscore SN$_{0}$ & $\;\;\; 0.003742$ & $-0.504857$ & $\;\;\; 0.003014$ & $\;\;\; 1.016900$ & $-0.185422$\\
Train\textunderscore SN$_{1}$ & $\;\;\; 0.012838$ & $\;\;\; 0.135103$ & $\;\;\; 0.013745$ & $\;\;\; 0.572341$ & $\;\;\; 0.584960$\\
Train\textunderscore SN$_{2}$ & $-0.024537$ & $\;\;\; 0.123520$ & $-0.024508$ & $\;\;\; 0.652294$ & $\;\;\; 0.252721$\\
Train\textunderscore SN$_{3}$ & $\;\;\; 0.004242$ & $\;\;\; 0.563471$ & $\;\;\; 0.004089$ & $-0.021505$ & $\;\;\; 0.326422$\\
\enddata
\tablecomments{Table~\ref{tab:allcoeffs} is published in its entirety in the machine-readable format. The first few rows and columns are provided here for guidance regarding its format.}
\end{deluxetable}

\begin{figure}
\includegraphics[width=\textwidth]{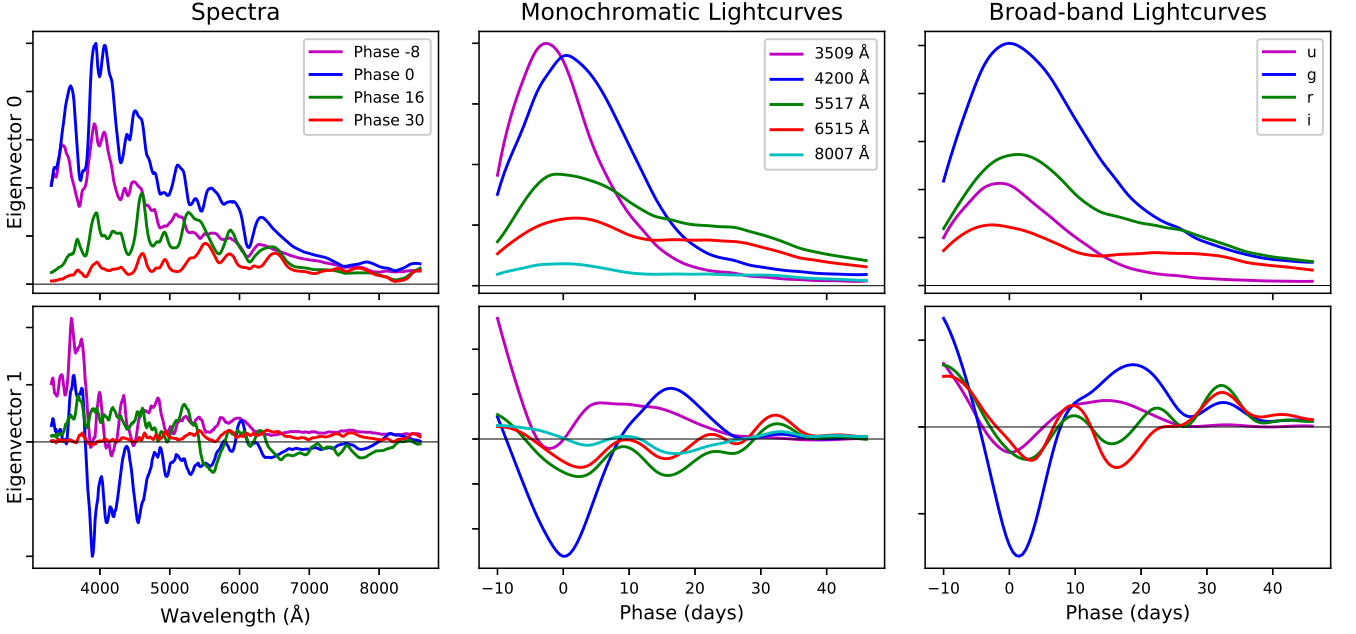}
\caption{Demonstrative spectra and lightcurves produced using \textsc{SNEMO2}. The left column shows slices in the phase dimension of the spectral time series (i.e., spectra) at four demonstrative phases, while the middle column shows slices in the wavelength dimension (monochromatic lightcurves). The right column shows broad-band lightcurves in the SDSS $u$, $g$, $r$ and $i$ filters.} 
\label{fig: Model-2}
\end{figure}

\begin{figure}
\includegraphics[width=\textwidth]{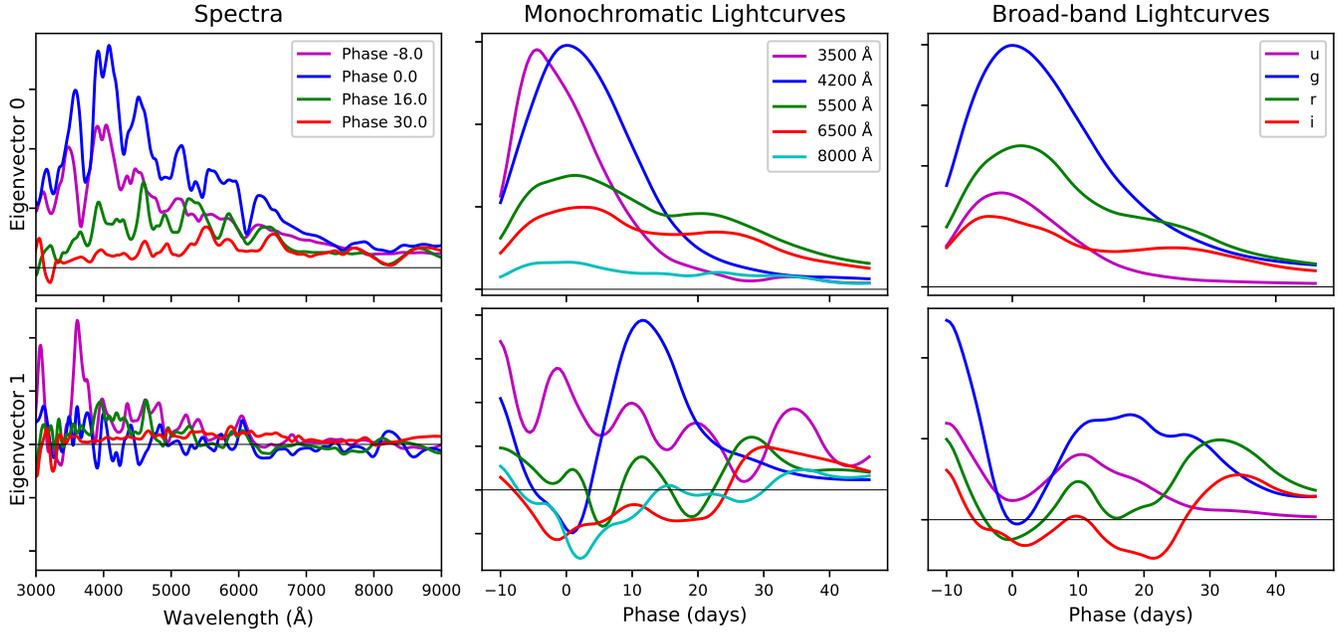}
\caption{Demonstrative spectra and lightcurves produced using the \textsc{SALT2} Model. See Figure~\ref{fig: Model-2} for descriptions of each panel.} 
\label{fig: Model-salt}
\end{figure}

\begin{figure}
\includegraphics[width=0.8\textwidth]{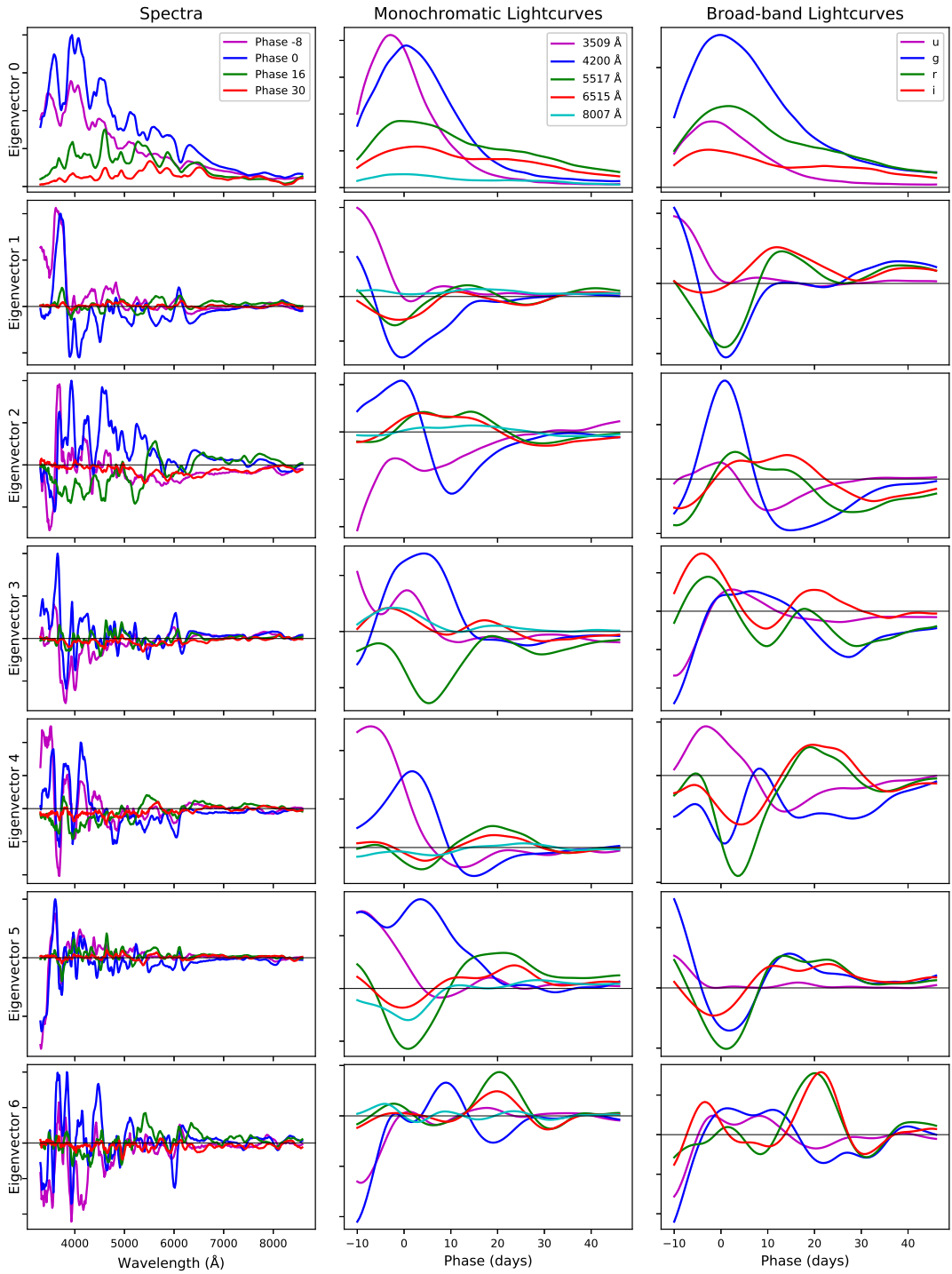}
\centering
\caption{Demonstrative spectra and lightcurves produced using \textsc{SNEMO7}. See Figure~\ref{fig: Model-2} for descriptions of each panel.} 
\label{fig: Model-7}
\end{figure}

\begin{figure}
\includegraphics[width=0.8\textwidth]{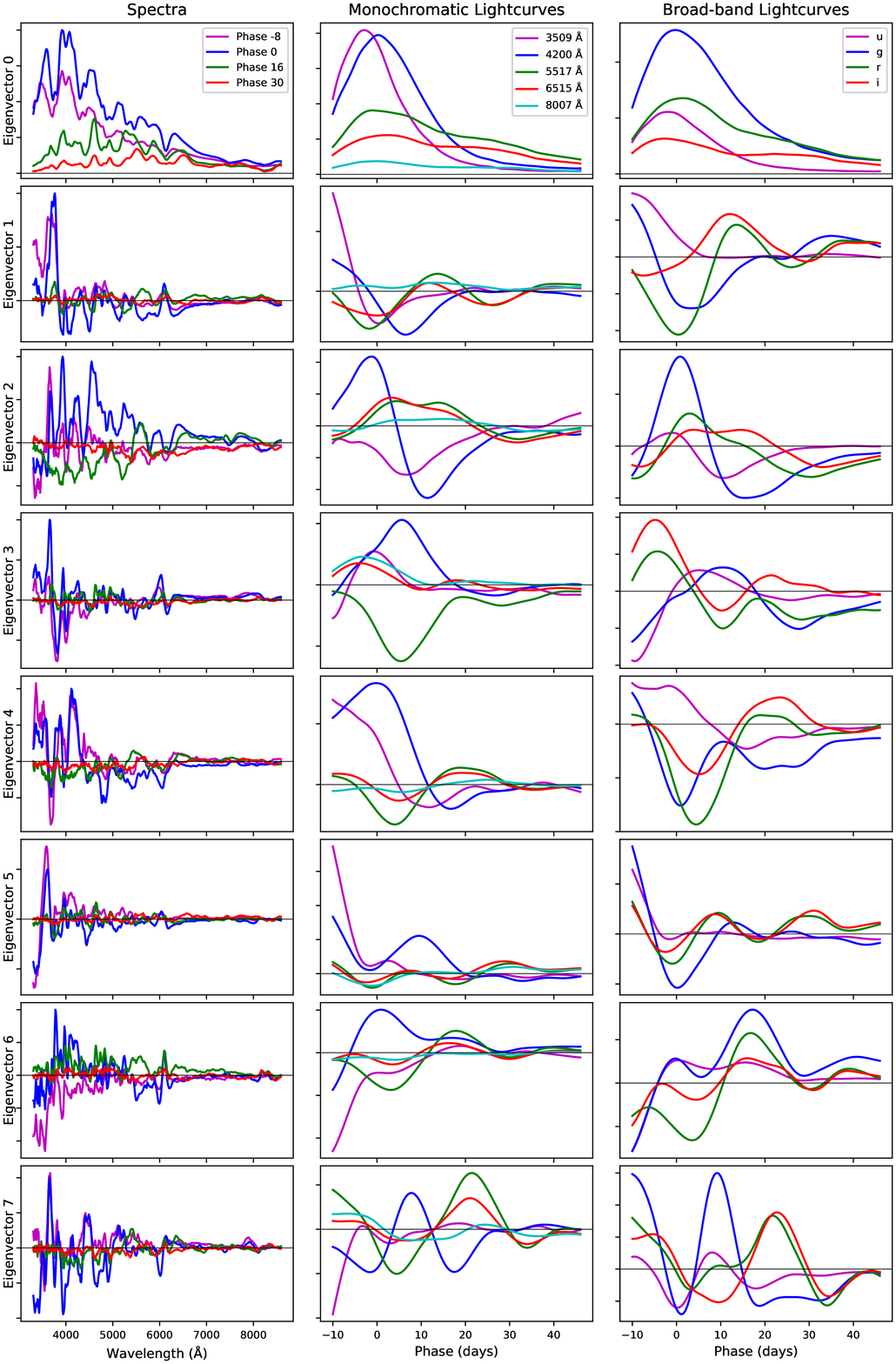}
\centering
\caption{Demonstrative spectra and lightcurves produced using the first eight components of \textsc{SNEMO15}. See Figure~\ref{fig: Model-2} for descriptions of each panel.} 
\label{fig: Model-15}
\end{figure}

The components of the models are shown in Figures~\ref{fig: Model-2}, \ref{fig: Model-7}, and \ref{fig: Model-15} (Figure~\ref{fig: Model-salt} shows the \textsc{SALT2} model for comparison). For demonstration purposes, we only show spectra at a few intervals in phase and monochromatic lightcurves at illustrative wavelengths. Another version of the models is shown in the Appendix, in which no outlier supernovae (which here were all identified after the fact as known SN1991T or SN1991bg-like objects) are included in the analysis. In these no-outlier training versions, about $1$ less component is needed for the \textsc{SNEMO7} and \textsc{SNEMO15}-analogue models.


\begin{figure}
\includegraphics[width=\textwidth]{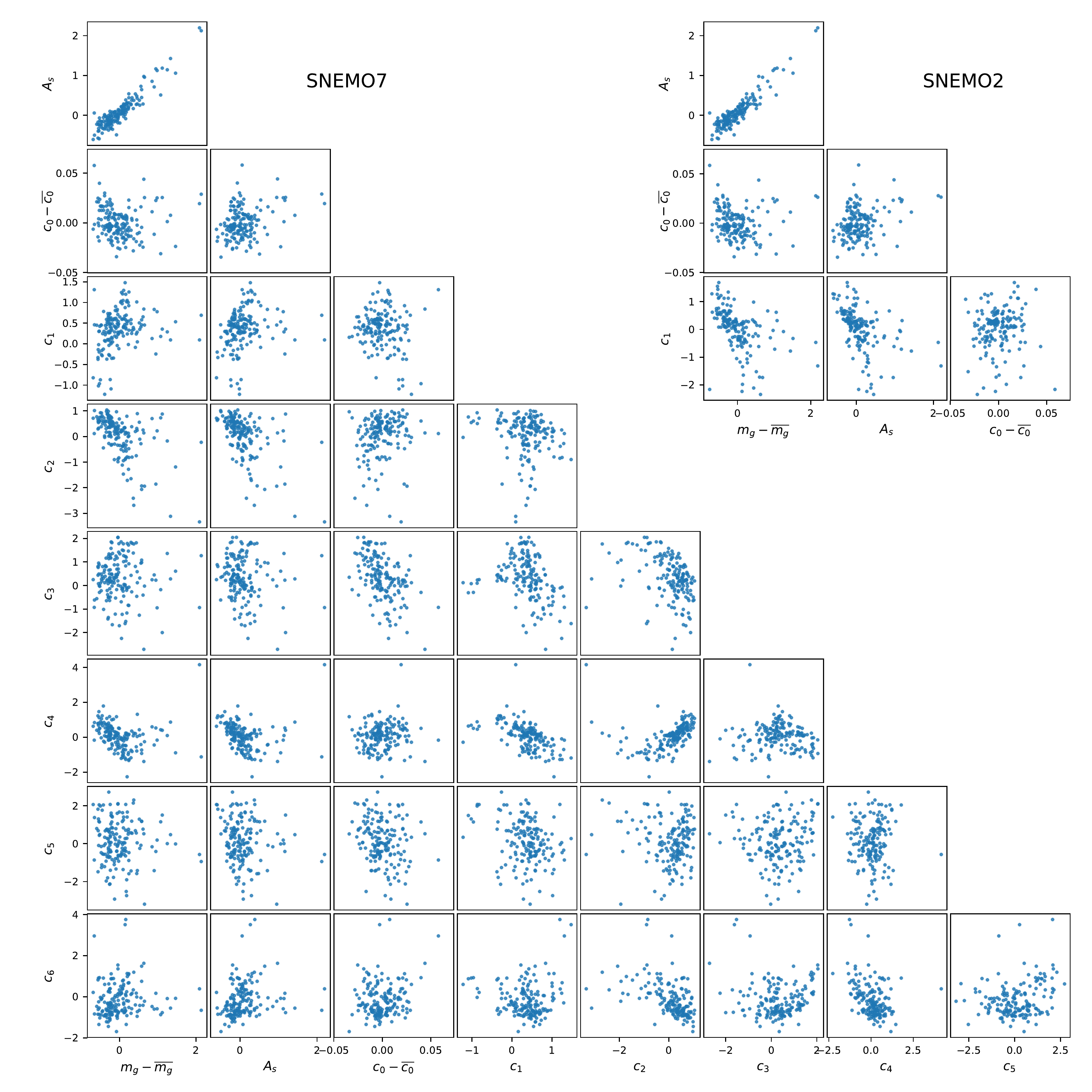}
\caption{Coefficient distributions for \textsc{SNEMO2} (small triangle in top right) and \textsc{SNEMO7} (large lower triangle).}
\label{fig: Coeffs}
\end{figure}

The distributions of coefficients for \textsc{SNEMO2} and \textsc{SNEMO7} are shown in Figure~\ref{fig: Coeffs}.
As expected, the $A_s$ parameter correlates strongly with $m_g$. The distributions provide hints of non-linear structure in the supernova population, but do not give any immediate indications of subpopulations.

It would be expected that the more complex models contain all of the information in the less complex models. The following matrix equations demonstrate how the \textsc{SNEMO2} and \textsc{SNEMO7} models are reconstructed by the more complex models, with the $\mathbf{R^2}$ values denoting coefficients of determination, i.e. the fraction of the variance explained in the model reconstruction of the other model (with $1$ as a perfect reconstruction).

For \textsc{SNEMO2}, 
\renewcommand*{\arraystretch}{.6}
\begin{equation} 
\label{eq:SNEMO2}
\mathbf{e^{(2)}} \approx 
 \mathbf{e^{(7)}} \cdot
\begin{pmatrix} 
\begin{array}{rr}
0.996 & 0.001 \\ 
0.515 & -0.612 \\ 
0.153 & 0.450 \\ 
0.080 & 0.067 \\ 
0.118 & 0.282 \\ 
0.254 & 0.089 \\ 
0.195 & -0.588
\end{array}
 \end{pmatrix} 
 \approx
  \mathbf{e^{(15)}} \cdot
\begin{pmatrix} 
\begin{array}{rr}
0.990 & 0.001 \\ 
0.546 & -0.492 \\ 
0.119 & 0.479 \\ 
0.103 & 0.210 \\ 
-0.173 & -0.253 \\ 
0.119 & -0.348 \\ 
-0.122 & 0.591 \\ 
-0.038 & 0.168 \\ 
-0.354 & 0.456 \\ 
0.204 & -0.285 \\ 
0.129 & -0.074 \\ 
-0.087 & 0.053 \\ 
0.160 & 0.047 \\ 
-0.322 & 0.157 \\ 
-0.178 & -0.240
\end{array}
 \end{pmatrix} 
\end{equation}
with corresponding $\mathbf{R^2}$ values
\begin{equation}
\mathbf{R^2_{(SNEMO7)}} = \begin{pmatrix} 1.000, & 0.857 \end{pmatrix} \quad \quad\mathbf{R^2_{(SNEMO15)}} = \begin{pmatrix} 1.000, & 0.871 \end{pmatrix}
\end{equation}
While for \textsc{SNEMO7},
\begin{equation}
\label{eq:SNEMO7}
\mathbf{e^{(7)}} \approx
\mathbf{e^{(15)}} \cdot
\begin{pmatrix} 
\begin{array}{rrrrrrr}
0.994 & 0.000 & -0.000 & -0.000 & -0.000 & 0.000 & -0.000 \\ 
0.039 & 0.952 & 0.158 & 0.008 & 0.060 & -0.055 & 0.012 \\ 
0.003 & -0.077 & 0.951 & -0.064 & 0.021 & 0.037 & -0.009 \\ 
0.008 & -0.038 & 0.108 & 0.999 & 0.132 & 0.055 & -0.025 \\ 
-0.093 & 0.108 & 0.023 & 0.088 & -0.844 & -0.082 & -0.045 \\ 
-0.203 & 0.255 & -0.269 & -0.092 & -0.211 & 0.858 & 0.124 \\ 
-0.018 & -0.257 & 0.317 & -0.090 & -0.047 & 0.496 & -0.262 \\ 
0.189 & -0.432 & 0.190 & -0.051 & -0.088 & -0.232 & 0.678 \\ 
-0.157 & 0.088 & -0.159 & -0.308 & 0.401 & 0.008 & -0.422 \\ 
0.019 & 0.249 & -0.155 & -0.060 & 0.608 & 0.204 & 0.295 \\ 
-0.059 & 0.119 & 0.135 & 0.077 & -0.158 & 0.129 & 0.128 \\ 
-0.064 & -0.089 & -0.227 & -0.245 & 0.226 & -0.181 & -0.226 \\ 
0.143 & 0.145 & -0.235 & 0.084 & 0.163 & 0.238 & -0.081 \\ 
-0.386 & 0.251 & 0.183 & -0.100 & 0.541 & -0.261 & -0.167 \\ 
-0.461 & 0.045 & -0.278 & 0.172 & 0.410 & -0.214 & -0.018
 \end{array}
\end{pmatrix} 
\end{equation}
with corresponding $\mathbf{R^2}$
\begin{equation}
\mathbf{R^2} = \begin{pmatrix} 1.000, & 0.988, & 0.968, & 0.984, & 0.916, & 0.986, & 0.900 \end{pmatrix}
\end{equation}

\renewcommand*{\arraystretch}{1.}
\normalsize
The somewhat lower $\mathbf{R^2}$ values for the second component of \textsc{SNEMO2} and for the seventh component of \textsc{SNEMO7} indicate small differences between the subspaces of the models.

\subsection{Overview of the Three Models}
\begin{figure}
\includegraphics[width=\textwidth]{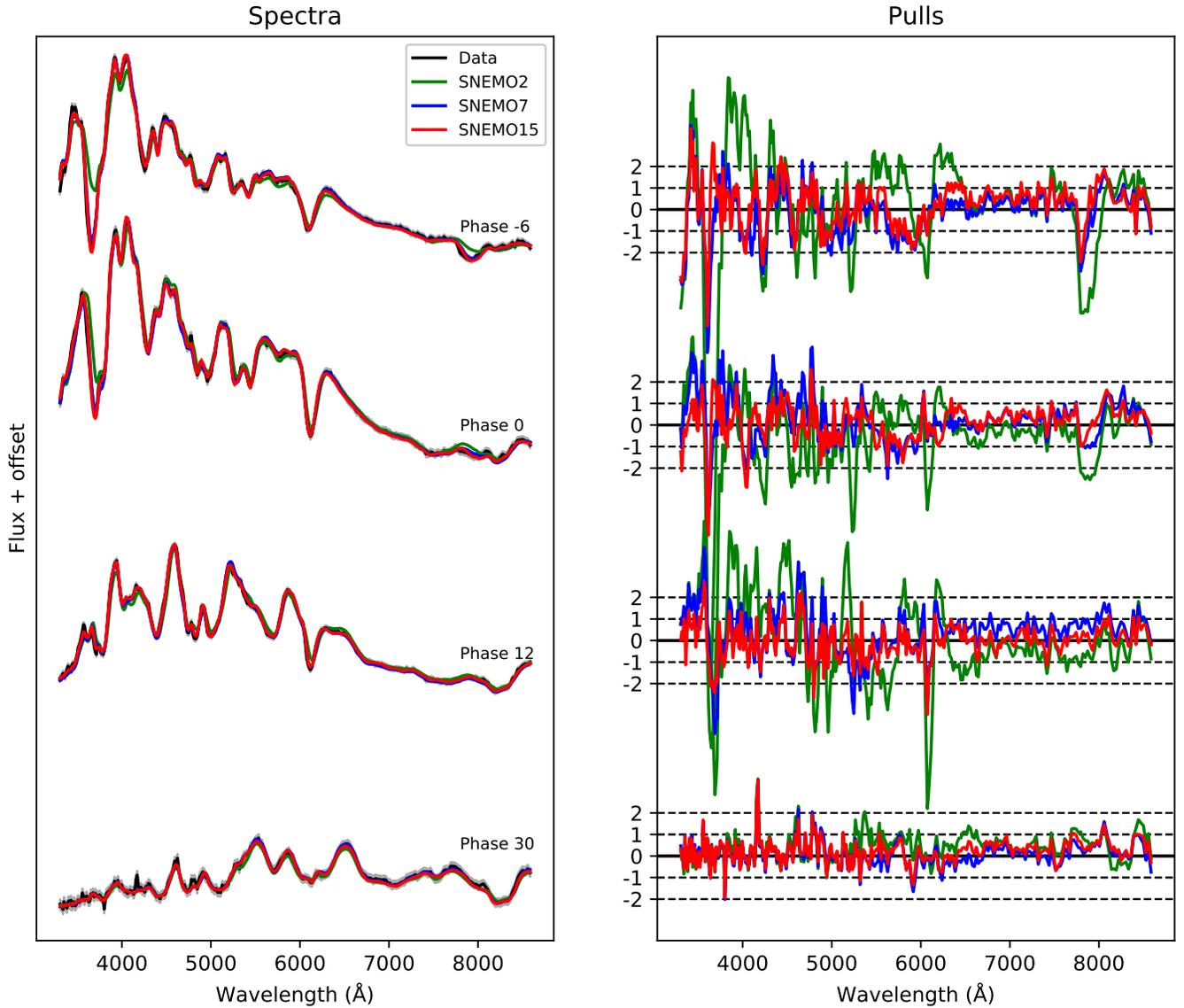}
\caption{Model reconstructions of an out-of-sample supernova. The supernova spectra and models for four demonstrative phases are shown in the left panel, while the right panel shows the pulls (residual divided by the uncertainty) of the models.} 
\label{fig: reconstruction}
\end{figure}

Figure~\ref{fig: reconstruction} shows an example of fitting a supernova with each of the three models. Here and after, the \textsc{SNEMO} models have been fit to the Gaussian Processes models of the hold-out supernovae, but the models can be equally applied to raw spectra or photometry by interpolating or integrating over the spectral time series templates. It can be seen from the residuals between the supernova spectra and the models that there are major improvements made moving from the two-component model, \textsc{SNEMO2}, to the seven-component model, \textsc{SNEMO7}. In the example shown, \textsc{SNEMO2} misses both broad-band behavior and spectral features of the supernova. Most of these are matched by \textsc{SNEMO7}. \textsc{SNEMO15} makes some additional improvements, particularly with adjustments to detailed shapes of the spectral feature Ca~II~H\&K and the early spectra.

\begin{figure}
\includegraphics[width=\textwidth]{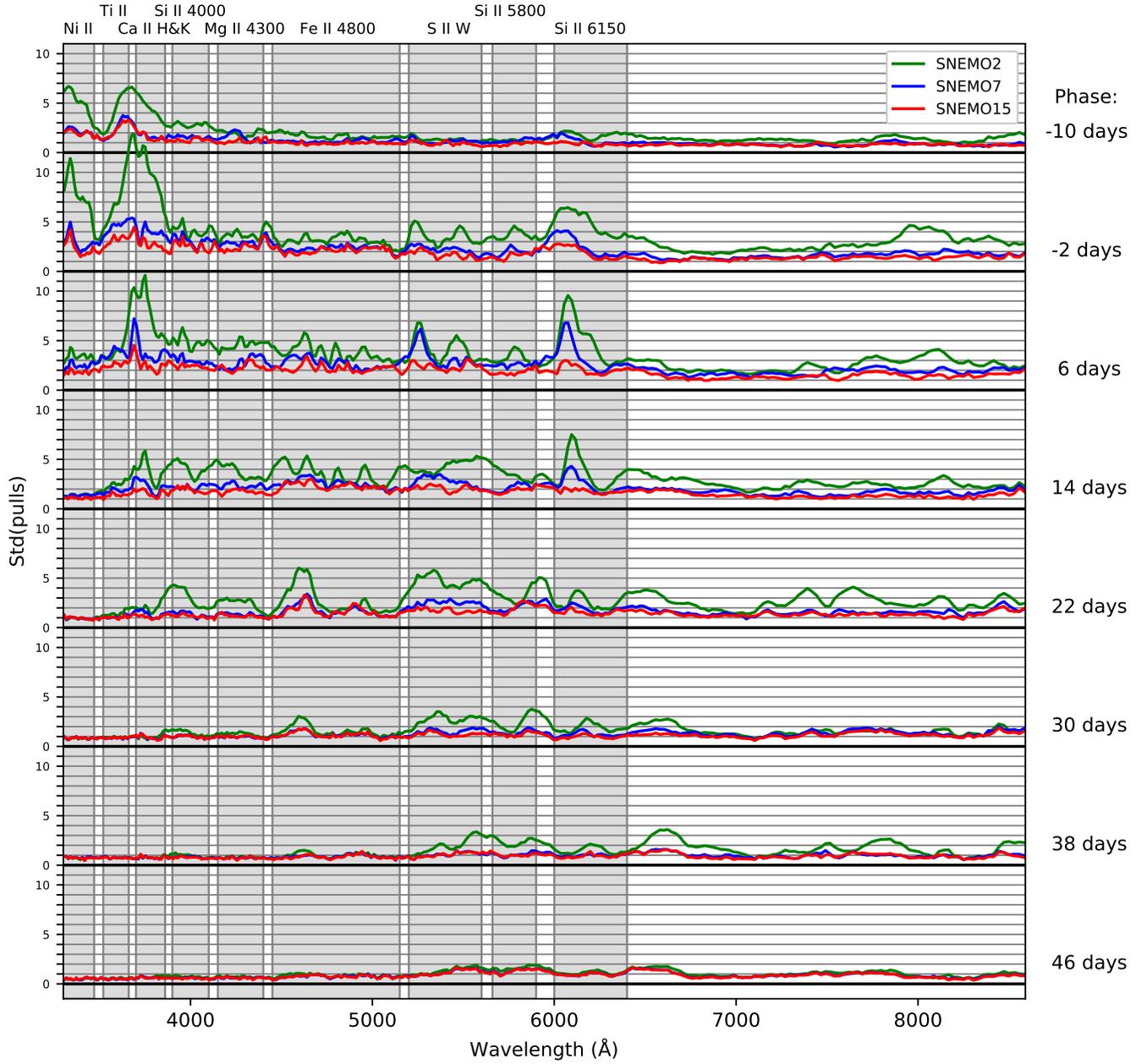}
\caption{Standard deviation of the pulls between the three \textsc{SNEMO} models and the supernovae from the hold-out set at demonstrative phases, indicated at the right of the figure. The $std(\mathrm{pulls})=1$ level indicates the level at which the residuals are on the same scale as the uncertainty in the data. The approximate regions of spectral features that strongly affect supernova spectra around maximum light are indicated by the vertical gray bands.}
\label{fig: std resids}
\end{figure}

A comprehensive view is found looking at the standard deviation of the residuals for all of the supernovae. This is shown at intervals in phase in Figure~\ref{fig: std resids}. Areas with large standard deviations in the residuals indicate spectral diversity that is not captured by the model. For \textsc{SNEMO2}, there are large residuals at the Si~II~6150 feature, with smaller residuals at the Si~II features near $4100 \mathrm{\AA}$. Large residuals are also apparent in the near-UV and $\mathrm{Ca \; H\&K}$ region, particularly around maximum light, and in the Fe II blend around $5000\mathrm{\AA}$ at and after maximum light. Aside from spectral features in the residuals, broad-band residuals are also apparent at and after maximum light. 

For \textsc{SNEMO7}, the spectral and broad-band residuals are greatly reduced. Some unmodelled supernova diversity is still apparent in the S II features, at Si II 6150 and in the near-UV near maximum light. Moving further to \textsc{SNEMO15}, almost all of these are gone and the residuals are at or below the scale of the error in the data. 

Table~\ref{table: broadband} shows the effect of adding or subtracting the one-sigma contribution (as determined by the distribution of model coefficients) of a given model component from the mean spectrum, i.e. the difference between certain characteristics of $e_{0}$ versus $e_{0} \, \pm \, std (c_{i}) \, e_{i}$. We consider the $g - r$ color at maximum and  the change in the $g$-band magnitude between $0$ and $15$ days after maximum light, $\Delta \, m 15(g)$ (thus the term here refers to the quantity itself, not a model fit parameter as for example in SNooPy, \citealt{Burns:2010vn}). These quantities are calculated by evaluating the broad-band synthetic photometry of the model at $0$ and $15$ days. In terms of broad-band behavior, \textsc{SNEMO2} is shown to be analogous to the \textsc{SALT2} model, with Eigenvector~$1$ causing effects of a similar magnitude in $\Delta \, m 15(g)$, though the effect on the $g - r$ color at maximum is larger for \textsc{SNEMO2}. The components can also have an effect on the date of maximum. Here we show the epoch of the $g$-band maximum, based on an interpolation of the components. The $g$-band maximum is slightly earlier than the phase zeropoint given by \textsc{SALT2} for all of the models (as well as for \textsc{SALT2}). For the reconstructed supernova models, there is a small scatter of $0.3$ days in the date of maximum, which is in agreement with the uncertainty in the \textsc{SALT2} phases used as input. 

In accordance with FM07, a constant value of $R_{V}=3.1$ has been assumed in this analysis. However, it can be seen in Table~\ref{table: broadband} that the components affect the $g - r$ color at maximum and, if the training supernovae have been affected by variable $R_{V}$, the model components might be have some capacity for fitting other values of $R_{V}$. This possibility was tested by fitting the models to supernova spectral time series that were simulated with higher and lower values of $R_{V}$, with a range of values for $A_{V}$. It was found that the \textsc{SNEMO7} and \textsc{SNEMO15} can account for a portion of the variation due to a change in $R_{V}$ (a roughly $25\%$ improvement in the $\chi^2$ values of fits to supernova spectra simulated using values of $R_V$ higher or lower than $3.1$), but do not have the power to fully account for variable $R_{V}$.

\startlongtable
\begin{deluxetable}{lrrr}
\tabletypesize{\footnotesize}
\tablewidth{0pc}
\tablecolumns{4}
\tablecaption{Broadband Characteristics of the Eigenvectors\label{table: broadband}}
\tablehead{
\colhead{Eigenvector}  \vspace{-0.3cm} & \colhead{$g-r$} & \colhead{$\Delta$ m15} & \colhead{$\Delta \mathrm{t}_{max}$}  \\
 & \colhead{(mag)} & (mag) & (days)}
\startdata
\textbf{SALT2} &&\\
Mean Spectrum & $-0.208$ & $-0.918$ & $-0.190$\\
Eigenvector 1 & $^{-0.008}_{+0.008}$ & $^{+0.074}_{-0.079}$ & $^{-0.100}_{+0.080}$\\
\textbf{SNEMO2} &&\\
Mean Spectrum & $-0.299$ & $-0.921$ & $-0.26$\\
Eigenvector 1 & $^{+0.042}_{-0.039}$ & $^{+0.147}_{-0.150}$ & $^{-0.77}_{+0.35}$\\
\textbf{SNEMO7} &&\\
Mean Spectrum & $-0.318$ & $-0.927$ & $-0.44$\\
Eigenvector 1 & $^{-0.029}_{+0.025}$ & $^{+0.053}_{-0.050}$ & $^{-0.50}_{+0.29}$\\
Eigenvector 2 & $^{+0.051}_{-0.046}$ & $^{+0.189}_{-0.197}$ & $^{-0.48}_{+0.22}$\\
Eigenvector 3 & $^{-0.007}_{+0.007}$ & $^{+0.003}_{-0.003}$ & $^{-0.19}_{+0.17}$\\
Eigenvector 4 & $^{+0.021}_{-0.021}$ & $^{+0.028}_{-0.030}$ & $^{+0.06}_{-0.07}$\\
Eigenvector 5 & $^{-0.018}_{+0.018}$ & $^{+0.020}_{-0.020}$ & $^{-0.07}_{+0.07}$\\
Eigenvector 6 & $^{+0.004}_{-0.004}$ & $^{+0.009}_{-0.009}$ & $^{-0.05}_{+0.05}$\\
\textbf{SNEMO15} &&\\
Mean Spectrum & $-0.330$ & $-0.933$ & $-0.53$\\
Eigenvector 1 & $^{-0.065}_{+0.058}$ & $^{+0.012}_{-0.011}$ & $^{-0.35}_{+0.28}$\\
Eigenvector 2 & $^{+0.036}_{-0.033}$ & $^{+0.194}_{-0.204}$ & $^{-0.40}_{+0.20}$\\
Eigenvector 3 & $^{+0.012}_{-0.012}$ & $^{+0.016}_{-0.017}$ & $^{+0.17}_{-0.18}$\\
Eigenvector 4 & $^{+0.022}_{-0.023}$ & $^{+0.018}_{-0.018}$ & $^{+0.09}_{-0.10}$\\
Eigenvector 5 & $^{-0.001}_{+0.001}$ & $^{+0.023}_{-0.022}$ & $^{-0.07}_{+0.06}$\\
Eigenvector 6 & $^{-0.032}_{+0.031}$ & $^{+0.020}_{-0.020}$ & $^{+0.05}_{-0.05}$\\
Eigenvector 7 & $^{-0.005}_{+0.005}$ & $^{+0.002}_{-0.002}$ & $^{+0.08}_{-0.09}$\\
Eigenvector 8 & $^{-0.007}_{+0.007}$ & $^{+0.015}_{-0.016}$ & $^{-0.09}_{+0.08}$\\
Eigenvector 9 & $^{-0.014}_{+0.014}$ & $^{+0.000}_{-0.000}$ & $^{+0.13}_{-0.13}$\\
Eigenvector 10 & $^{+0.001}_{-0.001}$ & $^{+0.003}_{-0.003}$ & $^{-0.06}_{+0.07}$\\
Eigenvector 11 & $^{-0.005}_{+0.005}$ & $^{+0.006}_{-0.006}$ & $^{+0.00}_{+0.01}$\\
Eigenvector 12 & $^{-0.011}_{+0.011}$ & $^{+0.009}_{-0.009}$ & $^{+0.01}_{+0.00}$\\
Eigenvector 13 & $^{+0.002}_{-0.002}$ & $^{+0.000}_{-0.000}$ & $^{-0.10}_{+0.11}$\\
Eigenvector 14 & $^{+0.006}_{-0.006}$ & $^{+0.002}_{-0.002}$ & $^{+0.02}_{-0.02}$\\
\enddata
\end{deluxetable}

\subsection{Comparison of \textsc{SNEMO2} with \textsc{SALT2}}
Comparing the two-component plus color model presented here with the \textsc{SALT2} model, Figures~\ref{fig: Model-2} and \ref{fig: Model-salt}, can give some context for understanding the new model components. The \textsc{SALT2} auxiliary template, corresponding to the $x_1$ parameter, is known to be highly correlated with the width of the B-band supernova lightcurve, so comparing the new model components with this template gives us a proxy for how correlated our model components are with lightcurve width.

\begin{figure}
\includegraphics[width=\textwidth]{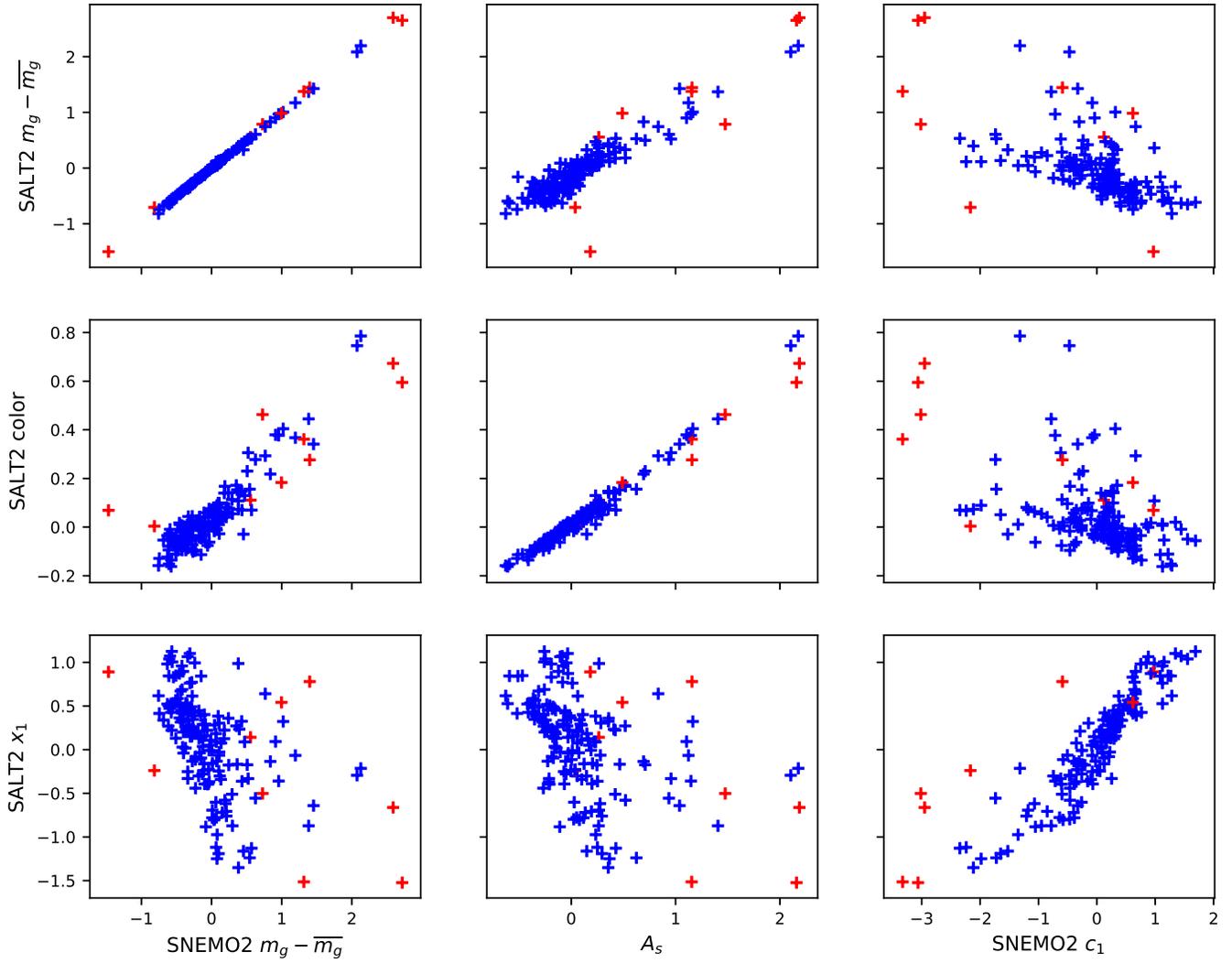}
\caption{Coefficients for \textsc{SNEMO2} versus the model coefficients for the \textsc{SALT2} model. Red points indicate supernovae that were identified as outliers.}
\label{fig: model2 vs SALT2}
\end{figure}

The supernova model parameters are compared in Figure~\ref{fig: model2 vs SALT2}. While the second components of each (Eigenvector~$1$) do not appear very similar, it is clear that the coefficients are highly correlated (bottom right panel of the figure). Thus, while they may be capturing the behavior differently, the two models find roughly the same way to discriminate among the supernovae. Some difference may be attributable to the fact that the \textsc{SALT2} $x_{1}$ and $c$ components are constructed to be uncorrelated, while this is not the case with the models shown here. A comparison of \textsc{SNEMO7} and \textsc{SNEMO15} (not shown) shows that each also has one or two components that have strong correlations with the \textsc{SALT2} $x_{1}$ coefficient, though not so clearly as in \textsc{SNEMO2}.

As in Equations~\ref{eq:SNEMO2} and \ref{eq:SNEMO7}, the SALT2 model can be reconstructed with the \textsc{SNEMO} models. In this case, it is also necessary to fit the color difference between the models. The best-fit reconstructions are given by
\renewcommand*{\arraystretch}{.6}
\begin{equation}
\mathbf{SALT2}
 \approx 
\begin{pmatrix}
\mathbf{e^{(2)}} \\
\mathrm{FM07}
\end{pmatrix} \cdot
\begin{pmatrix} 
\begin{array}{rr}
-0.440 & -0.012 \\ 
0.082 & -0.245 \\ 
0.817 & 0.388
\end{array}
 \end{pmatrix} 
 \approx
 \begin{pmatrix}
\mathbf{e^{(7)}} \\
\mathrm{FM07}
\end{pmatrix} \cdot
\begin{pmatrix} 
\begin{array}{rr}
-0.440 & -0.013 \\ 
-0.197 & 0.106 \\ 
-0.070 & -0.143 \\ 
-0.049 & 0.038 \\ 
0.020 & -0.230 \\ 
-0.014 & 0.218 \\ 
-0.238 & 0.166 \\ 
0.821 & 0.413
\end{array}
 \end{pmatrix} 
 \approx
 \begin{pmatrix}
\mathbf{e^{(15)}} \\
\mathrm{FM07}
\end{pmatrix}
\begin{pmatrix}
\begin{array}{rr}
-0.441 & -0.014 \\ 
-0.216 & 0.097 \\ 
-0.048 & -0.178 \\ 
-0.040 & -0.004 \\ 
0.011 & 0.208 \\ 
-0.057 & 0.374 \\ 
0.199 & 0.019 \\ 
-0.046 & -0.026 \\ 
0.339 & -0.216 \\ 
-0.150 & -0.078 \\ 
0.193 & 0.117 \\ 
0.140 & -0.222 \\ 
-0.035 & -0.049 \\ 
0.311 & -0.058 \\ 
-0.063 & 0.154 \\ 
0.828 & 0.515
\end{array}
 \end{pmatrix} 
 \end{equation}
\renewcommand*{\arraystretch}{1.}
\normalsize
where the first column refers to the \textsc{SALT2} mean template and the second column refers to the \textsc{SALT2} auxiliary template (which corresponds with the $x_1$ parameter). The corresponding $\mathbf{R^2}$ values are 
\begin{equation}
\mathbf{R^{2}_{(SNEMO2)}} = ( 0.994, 0.468) \quad
\mathbf{R^{2}_{(SNEMO7)}}  = (0.995, 0.673) \quad
\mathbf{R^{2}_{(SNEMO15)}} = (0.996, 0.694)
\end{equation}
The low value of $\mathbf{R^2}$ for the second component of \textsc{SALT2} is further indication of the differences between the  \textsc{SNEMO} models and \textsc{SALT2} on a spectral level, despite the correlations in the coefficient distributions in the supernova population.

\subsection{Minimizing Unexplained Magnitude Dispersion with \textsc{SNEMO7}}
\label{sec: disp final}
The seven-component model, \textsc{SNEMO7}, demonstrates the improvements made in linear standardization of magnitudes. As described in Section~\ref{sec: mag std}, we find a standardized magnitude by fitting a linear relation between the model coefficients and the $g$-band magnitude. The results are shown in Table~\ref{table: disp} using the terminology defined in Section~\ref{sec: mag std}. The parameters of the linear magnitude correction are calculated using the final hold-out set, since in a cosmological analysis the corrections would be recalculated for that given set of supernovae. For comparison and as a check for consistency, we also show the dispersion found when calculating the corrections using the training set plus the hold-out set. Note that while the Gaussian Processes step accounted for epoch-to-epoch flux calibration for an individual supernova, any calibration uncertainty between supernovae, aside from that due to peculiar velocity, is still part of the dispersion values found here.

Fitting the standardized magnitudes and the $\alpha$ and $\beta$ standardization parameters for the same supernovae using the \textsc{SALT2} model results in unexplained dispersions of $0.141^{+0.016}_{-0.021}$~mag ($0.149\pm0.019$~mag total dispersion) when fitting the hold-out set and $0.140^{+0.009}_{-0.010}$~mag ($0.150\pm0.009$~mag total dispersion) when fitting the training set plus the hold out set. For the \textsc{SALT2} model, $\sigma_{lc}\sim0.002$~mag, while $\sigma_{ext}$ is the same as for \textsc{SNEMO7}.

\begin{deluxetable}{l c c c c}
\tablecolumns{3}
\tablecaption{Final Results for the Dispersion in $g$-band Standardized Magnitudes using \textsc{SNEMO7}\label{table: disp}}
\tablehead{
\colhead{Data Set} \vspace{-0.2cm} & \colhead{Median $\sigma_{ext}$} & \colhead{Median $\sigma_{lc}$} & \colhead{ $\sigma_{u}$} & \colhead{ $\sigma_{tot}$} \\
& (mag) & (mag) & (mag) & (mag)
}
\startdata
Hold-out Set                                      & $0.041$ & $0.021$ & $0.121^{+0.023}_{-0.017}$ & $0.128\pm0.020$ \\ 
Hold-out Set with $3\sigma$ cuts & $0.041$ & $0.020$ & $0.087^{+0.021}_{-0.015}$ & $0.100\pm0.014$ \\
All SNe                                              & $0.045$ & $0.021$ & $0.110^{+0.010}_{-0.008}$ & $0.125\pm0.010$ \\ 
All SNe with $3 \sigma$ cuts        & $0.045$ & $0.021$ & $0.097^{+0.009}_{-0.008}$ & $0.113\pm0.007$ \\ 
\enddata
\label{table: disp}
\tablecomments{See Equation~\ref{eq: disp} for definition of terms.}
\end{deluxetable}

\begin{deluxetable}{l r}
\tablecolumns{2}
\tablecaption{Corrected Magnitude Best-fit Parameters\label{table: alphas}}
\tablehead{ \colhead{Component} & \colhead{Parameter}}
\startdata
$\alpha_c$ & $-1.081  \pm  0.040 $\\
$\alpha_1$  & $-0.074   \pm  0.014 $ \\
$\alpha_2$  &  $-0.018  \pm 0.021 $\\
$\alpha_3$  &  $-0.093  \pm  0.015 $\\
$\alpha_4$  & $-0.010 \pm  0.017 $\\
$\alpha_5$  & $-0.051  \pm 0.010 $\\
$\alpha_6$  &  $ 0.034  \pm  0.015$ \\
\enddata
\end{deluxetable}

The value for the unexplained dispersion is dependent on the values for $\sigma_{lc}$ and $\sigma_{ext}$, but is not highly sensitive to changes in their values. For example, if the uncertainties in $\sigma_{lc}$ were actually half of the value used here and reported in Table~\ref{table: disp}, the unexplained dispersion fit for all supernovae with $3 \sigma$ cuts would increase from $0.097$ to $0.103$ mag. Changing the assumed value of the host galaxy peculiar velocities used to calculate $\sigma_{ext}$ by $\pm100 \;\mathrm{km \; s^{-1}}$ changes the resulting unexplained dispersion by less than $\pm 0.01$ mag.

\begin{figure}
\includegraphics[width=\textwidth]{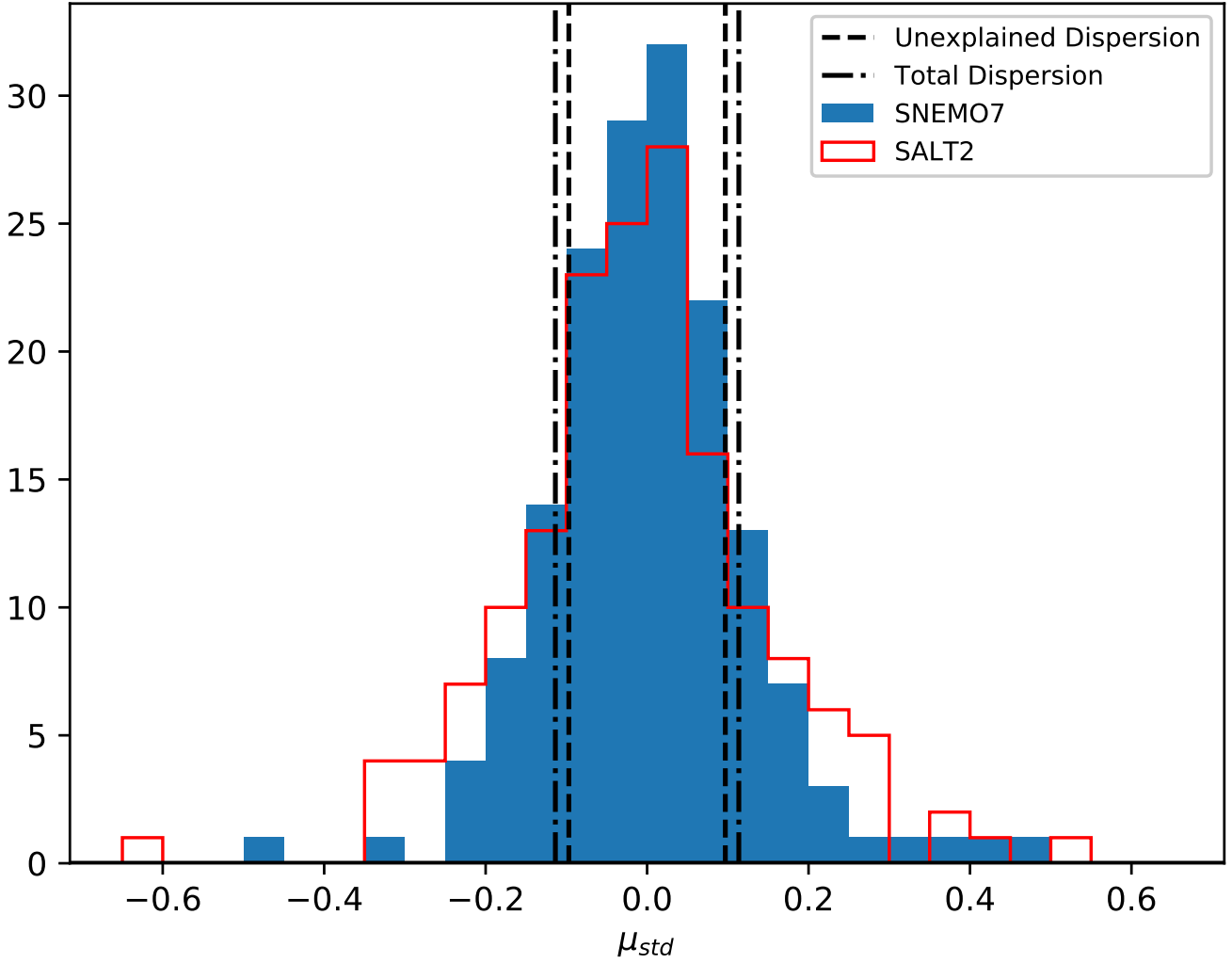}
\caption{Resulting $\mu_{std}$ for \textsc{SNEMO7} after minimizing Equation~\ref{eq: disp}. The \textsc{SALT2} $\mu_{std}$ for the same supernovae are showed as a comparison in red. The dashed lines show the level of the unexplained dispersion, while the dot-dashed lines show the total dispersion.}
\label{fig: final disp}
\end{figure}

As in the `Union' compilation (\cite{Amanullah:2010}), we remove supernovae with standardized magnitude pulls (the residual divided by the error) of more than $3$. This removes one supernova from the training set and one supernova from the hold-out set, which are not otherwise outliers in the coefficient space. Similarly, one of these supernovae was also determined to be an outlier and was removed for calculating the \textsc{SALT2} dispersion. The resulting dispersion is lower than the \textsc{SALT2} values but not quite at a level equal to that of `twin' supernovae (F15, discussed further below), which found an unexplained dispersion of $0.072\pm 0.010$ mag. (Note that the twinning technique makes no assumption about the linearity of the relation between the spectra or lightcurves and magnitudes.) This difference indicates that moving from linear standardization to a more sophisticated method may be necessary to maximize the possible improvement in standardized magnitudes. The standardized magnitude residuals are shown in Figure~\ref{fig: final disp}, compared to the residuals for the same supernovae using \textsc{SALT2}.

Table~\ref{table: alphas} shows the best fit parameters for the linear correlations between the model coefficients and the $g$-band magnitude, corresponding to the $\alpha_i$ parameters in Equation~\ref{eq: disp}. Those shown are the result of fitting to all of the data (the training plus hold out sets) to give the most significant result. It can be seen that some of the components have a stronger significance, while the others are within $1-2\sigma$ of zero. This indicates that some components do not have a strong effect on the $g$ band magnitudes at maximum, though they may still affect other features of the lightcurve. Whether more data would enable significant correlations to be fit between all of the components and the $g$-band magnitude remains to be seen.

\subsection{Measures of the Reconstruction of the Spectral Time Series}
\begin{figure}
\includegraphics[width=\textwidth]{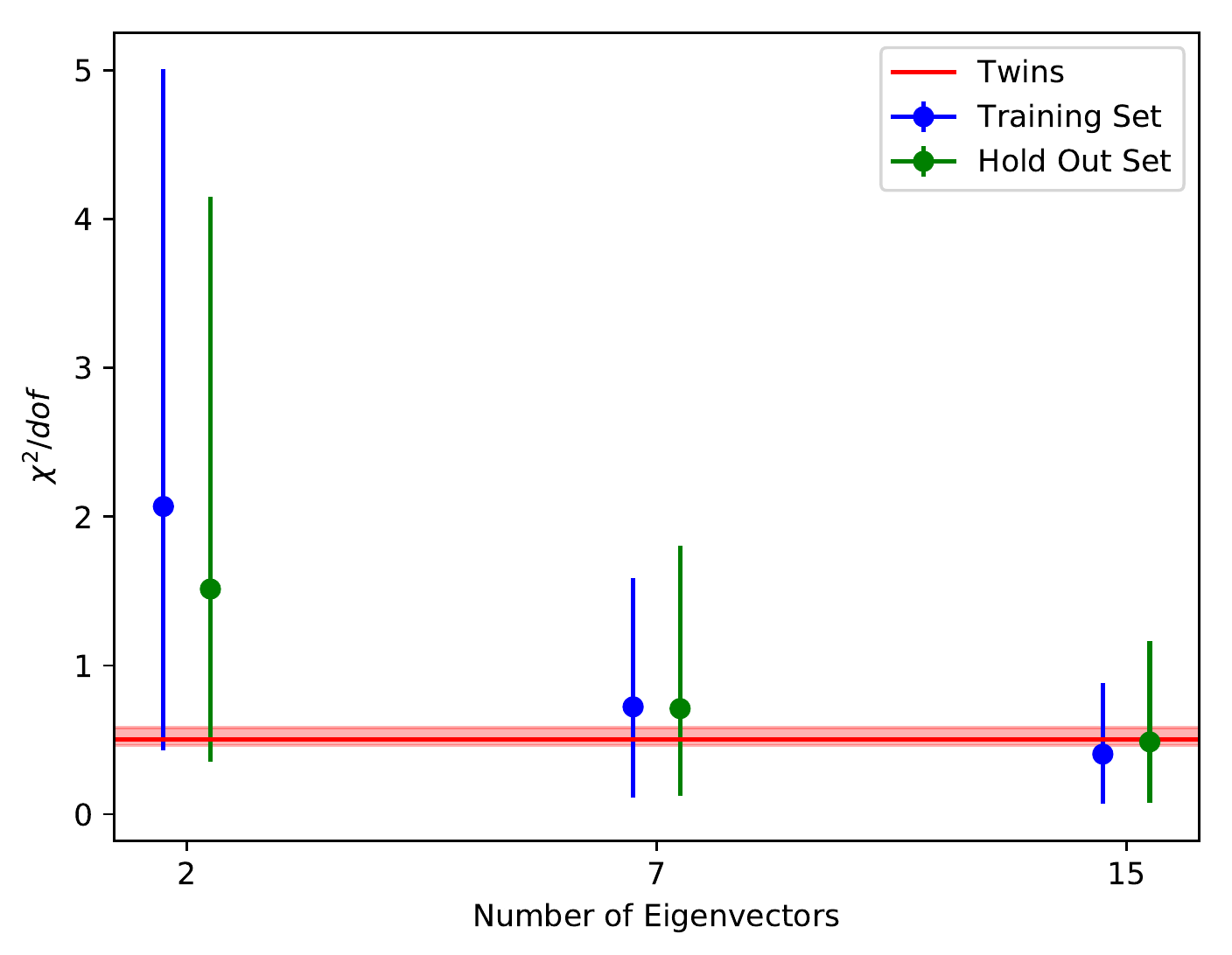}
\caption{Median and $95\%$ confidence interval for the reduced $\chi^2$ of the three models in comparison to the reduced $\chi^2$ of the best supernova twins. The blue points correspond to the result for the training set and the green points correspond to the smaller hold-out set supernovae. The red line is the median and $95\%$ confidence interval of the reduced $\chi^2$ between supernova twins. The corresponding reduced $\chi^2$ values for the \textsc{SALT2} model are $7.68^{+1.72}_{-1.90}$ for the training set and $6.59^{+2.23}_{-1.36}$ for the hold-out set.}
\label{fig: model vs twins}
\end{figure}

\begin{figure}
\includegraphics[width=\textwidth]{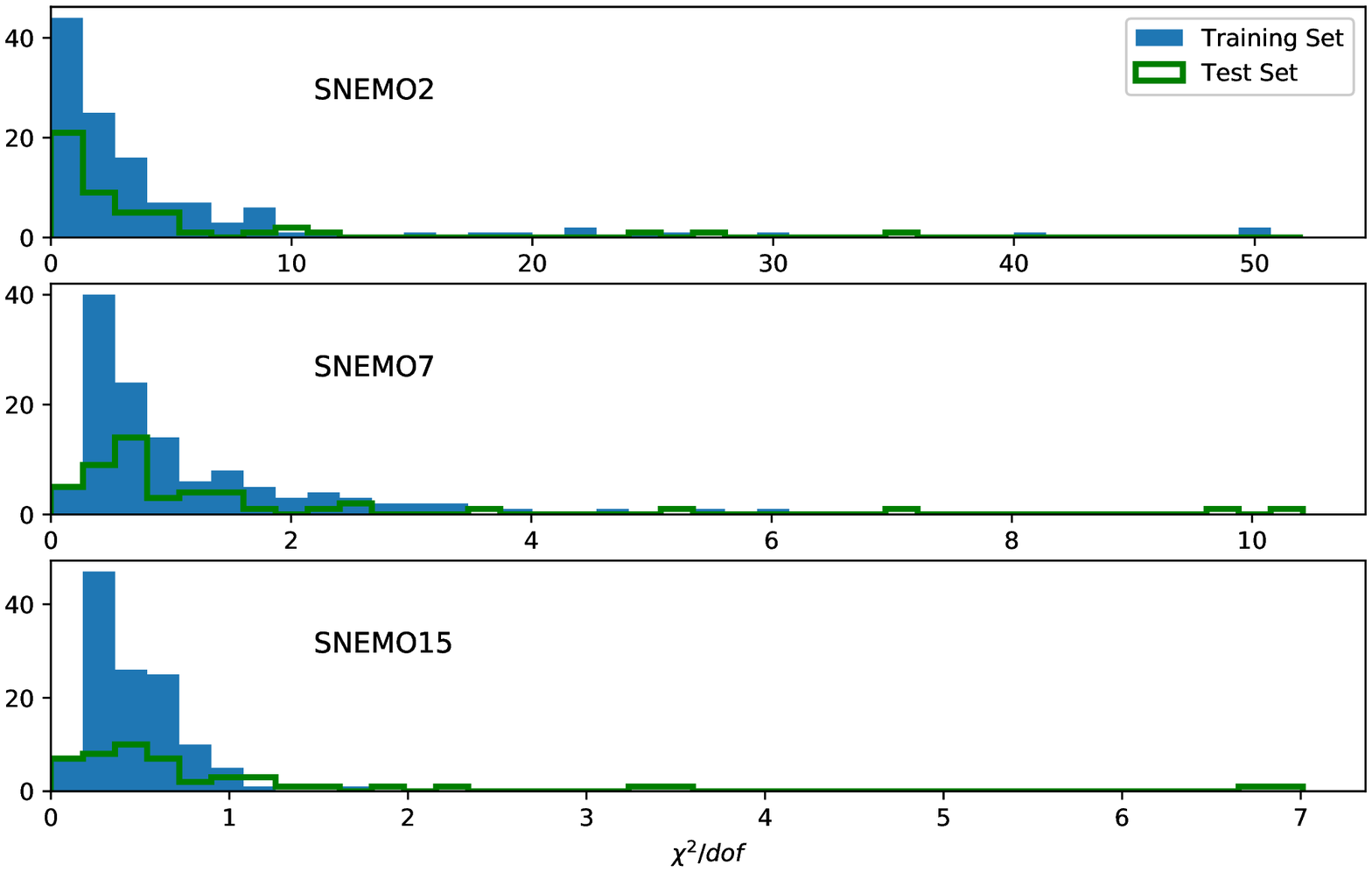}
\caption{The three panels show normalized histograms of the reduced $\chi^2$ of the supernovae for the three models in comparison. The histograms are normalized to more easily compare the training and hold-out set results. Note that the x-axis scales are different in each of the plots.}
\label{fig: model chis}
\end{figure}

Figure~\ref{fig: model vs twins} shows the median and $95\%$ confidence interval on the reduced $\chi^2$ values for \textsc{SNEMO2}, \textsc{SNEMO7}, and \textsc{SNEMO15} based on a bootstrap analysis of the training and hold-out set $\chi^2$ distributions. As discussed in Section~\ref{sec: Model Testing}, we are more concerned here with the relative values of the reduced $\chi^2$, rather than the absolute value, which is known to be underestimated. The differences in the sizes of the uncertainties are partially driven by the different sample sizes and also by outliers, which can be seen in the histograms in Figure~\ref{fig: model chis}. The values for the training and hold-out sets are consistent with the results found in the cross-validation process. Reduced $\chi^2$ values calculated in the same way for the \textsc{SALT2} model are $7.68^{+1.72}_{-1.90}$ for the training set and $6.59^{+2.23}_{-1.36}$ for the hold-out set.

Examination of the supernovae from the hold-out set that still have somewhat large reduced $\chi^2$ values when fit by \textsc{SNEMO15} (see Figure~\ref{fig: model chis}) shows that most of these are peculiar supernovae. It may be that from the small number of SN~1991T-like objects in the training set, we do not have enough information to fully capture the behavior for this type of supernova. One outlier is PTF09dnp, Test\textunderscore SN44 in Table~\ref{tab:allcoeffs}, an otherwise normal object that is a rare case of a supernova with very narrow absorption lines where the $\mathrm{Ca II\; H \& K}$ and $\mathrm{Si}$ lines can be separately resolved. It can be seen from the residual between the model and supernova spectra that this feature is driving the relatively large $\chi^2$, which is not unreasonable since this is the only supernova in our data set to have this feature. The quantity of such outlier $\chi^2$ values gives an indication of the scale at which we have captured the full diversity of Type Ia supernova spectral time series. It would be expected that an even larger sample of spectrophotometric time series would better incorporate this behavior, possibly requiring additional model components.

\subsubsection*{Comparison with `twin' Supernovae}
The reduced $\chi^2$ values for the \textsc{SNEMO} models can be compared with the analysis of `twin' supernovae from F15. In that analysis, pairs of supernovae were identified in the SNfactory data set that, after normalizing and dereddening, had very similar spectral time series. The concept underpinning that analysis was that such twin supernovae would have the same intrinsic luminosities and thus would be better standard candles requiring little or no correction to relative distance measurements. These supernovae were called twins, and it was found that there was a dispersion of $0.072$~mag among the magnitudes of twins, which is the best that has been found for a large sample of Type Ia supernovae.

Here, we took Gaussian Process predictions for the best matches from the supernova twins analysis and calculated the $\chi^2$ values, showing how closely the spectral time series for a pair of twins match each other (after normalizing and dereddening the spectra, as in the twins analysis). The median value for these is shown as the red line in Figure~\ref{fig: model vs twins}, where it is at about the level of the median reduced $\chi^2$ for \textsc{SNEMO15} fit to the hold-out set.

This shows that at fifteen components, our model is matching the supernova spectra at the same level that good twins match each other. The quality of the spectral matching implies that with \textsc{SNEMO15} it should be possible to achieve the same level of magnitude standardization as with twins. The fact that the dispersion in the standardized magnitudes is poorer than for twins, even though the reduced $\chi^2$ is similar, points to the linear standardization method as the limit in improving magnitude dispersion with \textsc{SNEMO}. Improvements could be made in two ways: one way would be to fit supernovae with the model and identify cases that are very close to one another in the space of model coefficients, effectively finding twins through the coefficients of the model instead of directly fitting pairs of spectra. As with twins, one would then use the `model twins' as standard candles. The other possibility is to take a more holistic approach and find a mapping from the space of model coefficients onto the possible range of $g$-band magnitudes. If the coefficients are a unique identifier of the supernova spectral time series, then there must be some holomorphic map from them to the magnitudes, either by means of an analytic function (for example the linear method used in the standardization here) or something like a machine learning technique. This hypothetical mapping would need to be able to follow nonlinear and non-orthogonal relationships between the model coefficients and the supernova magnitudes. In principle, this should achieve a level of magnitude dispersion comparable with the twins method, to the extent that it describes an equivalent amount of the spectral time series. However, such a mapping might not be discoverable without a larger data set than available here.

\section{Discussion and Conclusions}
\label{sec: Conclusion}
In sum, we have calculated three models for Type Ia supernovae intended for a range of applications. The two-component model, \textsc{SNEMO2}, serves mainly as a tool for comparison with other two-component empirical models for supernova spectral time series currently in use. Though the components of the model appear quite different from the components in the \textsc{SALT2} model, they describe similar effects in supernova behavior, which can be seen from the correlation in the model coefficients. 

The seven-component model, \textsc{SNEMO7}, describes much more of supernova diversity and is also the most efficient among the models presented here for making linear corrections to the g-band magnitude and minimizing dispersion among supernovae, assuming that the components can be accurately fit by the data. The number of model components is small enough that it should be possible to meaningfully constrain this model to supernova data with only photometric lightcurves, though testing of the necessary temporal sampling and signal-to-noise is left for future study.

Lastly, the fifteen-component model, \textsc{SNEMO15}, matches spectral time series for all the supernovae in our data set on the same level that the best `twin' supernovae match each other. This model is particularly suited to producing realistic supernova simulations. Given the number of components, it will probably be difficult to fit the parameters of this model without at least some knowledge of the spectral behavior of the supernova being fit. If the data can sufficiently constrain the model parameters, it should also be possible to achieve magnitude standardization at the level of supernova twins, around $0.07$ mag. A lightcurve fitter that could match supernova spectra at this level would also remove certain systematic errors, such as those due to $K$-corrections discussed in S15, which may not be the case with \textsc{SNEMO7}. This improvement will be possible if the model parameters can be fit accurately enough to minimize differences between the spectral-temporal model and the true supernova spectrum. If this accurate fitting could be done, the same standardized magnitude would be fit independent of the filters used, unlike what was found using current two-parameter lightcurve fitters in S15.

\medskip

Future work will determine how the models presented here can be best applied to observational data---whether external supernovae can be constrained by photometry alone or if spectral data will be required. Preliminary testing shows that the Si II velocity is recovered to within $700$ km s$^{-1}$ for \textsc{SNEMO7} and within $630$ km s$^{-1}$ for \textsc{SNEMO15} when the models are fit to spectral time series, and simulations will show how well these numbers hold up with photometric data. We can also begin to test more advanced nonlinear standardization techniques, moving beyond fitting linear correlations between the magnitudes and the model parameters, assuming that the number of supernovae available is sufficient to constrain such methods.

The models presented here will be incorporated into the supernova lightcurve fitting Python package \texttt{sncosmo} \footnote{https://sncosmo.readthedocs.io} (\citealt{Barbary:2014}), so that anyone can apply one of these models to their data. The coefficient distributions of the supernovae are published as part of the current paper so that the model can be used to create simulated data sets. 

As described in Section~\ref{sec: Data}, the hybrid standard star system used to calibrate the SNfactory data is in tension with data calibrated purely with the CALSPEC standards, an effect that must be accounted for in applying the current version of these models to new data. Since this calibration difference primarily affects the $U-B$ color, which is estimated to be $\sim0.02$~mag lower in the system used here, there may be small systematic biases if the models are applied to data calibrated to CALSPEC or other calibration systems if restframe $U$-band observations have a high weight in proportion to the rest of the data.However, work is in progress to allow the SNfactory data set to be recalibrated on a pure CALSPEC system, and when this is done the models presented here will be retrained to facilitate application to outside data sets (such as \citealt{Betoule:2014}) that are calibrated to the CALSPEC system.

Supernova cosmology has relied on two-component models for over twenty years. For most of this time, models were limited by the lack of sufficient data to make any improvements upon this. We now have the potential to move past this barrier, with advances possible both for our understanding of Type Ia supernovae and our capacity to measure dark energy.

\section*{Acknowledgments}
The authors are grateful to the technical and scientific staff of the University of Hawaii 2.2 m telescope, the Palomar Observatory, and the High Performance Wireless Radio Network (HPWREN). We thank Dan Birchall for his assistance in collecting data with SNIFS. We wish to recognize and acknowledge the very significant cultural role and reverence that the summit of Mauna Kea has always had within the indigenous Hawaiian community. This work was supported by the Director, Office of Science, Office of High Energy Physics, of the U.S. Department of Energy under Contract No. DE-AC02-05CH11231; by a grant from the Gordon \& Betty Moore Foundation; in France by support from CNRS/IN2P3, CNRS/INSU, and PNC; and in Germany by the DFG through TRR33 ``The Dark Universe." National Science Foundation Grant Number ANI-0087344 and the University of California, San Diego provided funding for HPWREN. In China support was provided from Tsinghua University 985 grant and NSFC grant No 11173017. CS was also supported by the Labex ILP (reference ANR-10-LABX-63) part of the Idex SUPER, and received financial state aid managed by the Agence Nationale de la Recherche, as part of the program ``Investissements d'avenir" under the reference ANR-11-IDEX-0004-02. This project has received funding from the European Research Council (ERC) under the European Union's Horizon 2020 research and innovation programme (grant agreement n°759194 - USNAC). Based in part on observations obtained with the Samuel Oschin Telescope and the 60 inch Telescope at the Palomar Observatory as part of the Palomar Transient Factory project, a scientific collaboration between the California Institute of Technology, Columbia University, Las Cumbres Observatory, the Lawrence Berkeley National Laboratory, the National Energy Research Scientific Computing Center, the University of Oxford, and the Weizmann Institute of Science.

\appendix

\label{app: out}
For calculating the correction function and the resulting dispersion in the standardized magnitudes we exclude extreme outlier supernovae. To do this we calculate the Mahalanobis distance (\citealt{Mahalanobis:1936}), or normalized metric distance, of each of the supernovae in the space of the model coefficients. The coefficient corresponding to the normalization of the model, $c_0$, which is degenerate with the supernova magnitude, is not included in this calculation. Supernovae are excluded that have a probability of less than $.001\%$ according to the expected $\chi^2$ distribution of the Mahalanobis distances. This procedure is then repeated once, since outliers affect the calculation of the Mahalanobis distances. This removed six supernovae from the standardization analysis. Three are 1991T-like supernovae, identified by the lack of $\mathrm{Si II}$ and $\mathrm{Ca II \; H\&K}$ absorption before maximum and strong absorption from $\mathrm{Fe II}$ and $\mathrm{Fe III}$ (\cite{Filippenko:1992b}) and three are 1991bg-like supernovae, mainly identified by broad $\mathrm{Ti II}$ absorption around $4200 \mathrm{\AA}$ (\cite{Filippenko:1992a}). One of the SN~1991T-like supernovae is the super-Chandrasekhar mass supernova SN~2007if (here called Train\textunderscore SN46), the SNfactory observations of which have been previously discussed in \cite{Scalzo:2010}.

Such supernovae, assuming they could be correctly identified, would generally be excluded in current cosmological analyses because they are not expected to be well fit by current lightcurve fitters like \textsc{SALT2} (\cite{Amanullah:2010}), though future non-linear methods for magnitude standardization may allow these types of supernovae to be brought back into the set used for cosmology. Since these are not currently useful for cosmology, there is some motivation to exclude them from the entire analysis: the model might be able to describe `normal' supernovae better or with fewer components if it was not required to also fit these peculiar supernovae. Testing showed that removing these supernovae does change the individual components of the models, but does not lead to a significant improvement in the $\chi^2$ of the model or the resulting dispersion in the standardized magnitudes, except in the two-component model, where it did improve the $\chi^2$ metric. K-fold cross validation showed that the favored model for standardizing magnitudes and the best spectral model were at six and fourteen components respectively, as opposed to seven and fifteen when the peculiar supernovae are included. Thus, there is the benefit of needing one less component to describe the more normal supernovae. We present the resulting models using two, six, and fourteen components in Figures~\ref{fig: model 2, no outs} through~\ref{fig: model 14, no outs, pt1}.

\begin{figure}
\includegraphics[width=\textwidth]{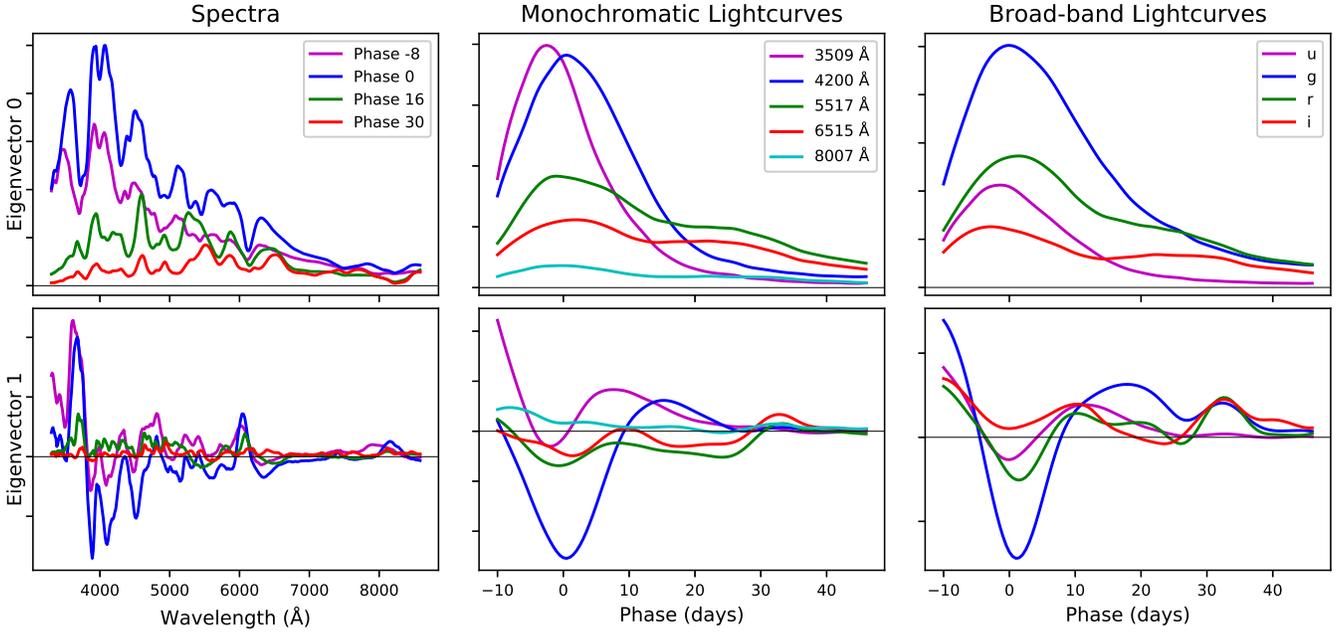}
\caption{Demonstrative spectra and lightcurves for the eigenvectors of the two-component model trained with no outlier supernovae. See Figure~\ref{fig: Model-2} for comparison and for description of the subplots.}
\label{fig: model 2, no outs}
\end{figure}

\begin{figure}
\includegraphics[width=\textwidth]{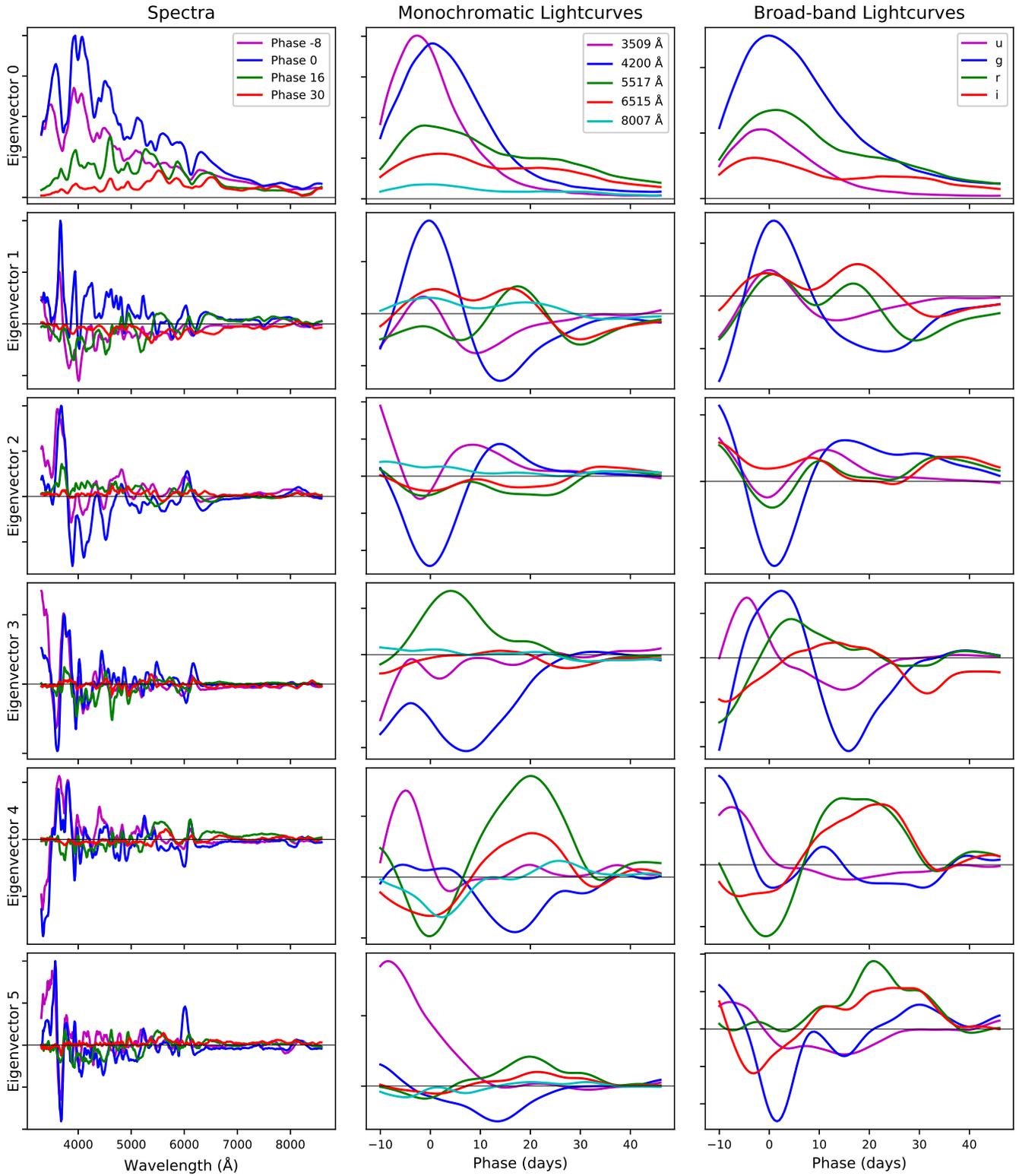}
\caption{Demonstrative spectra and lightcurves for the eigenvectors of the six-component model trained with no outlier supernovae. See Figure~\ref{fig: Model-2} for description of the columns.}
\label{fig: model 6, no outs}
\end{figure}

\begin{figure}
\includegraphics[height=\textheight]{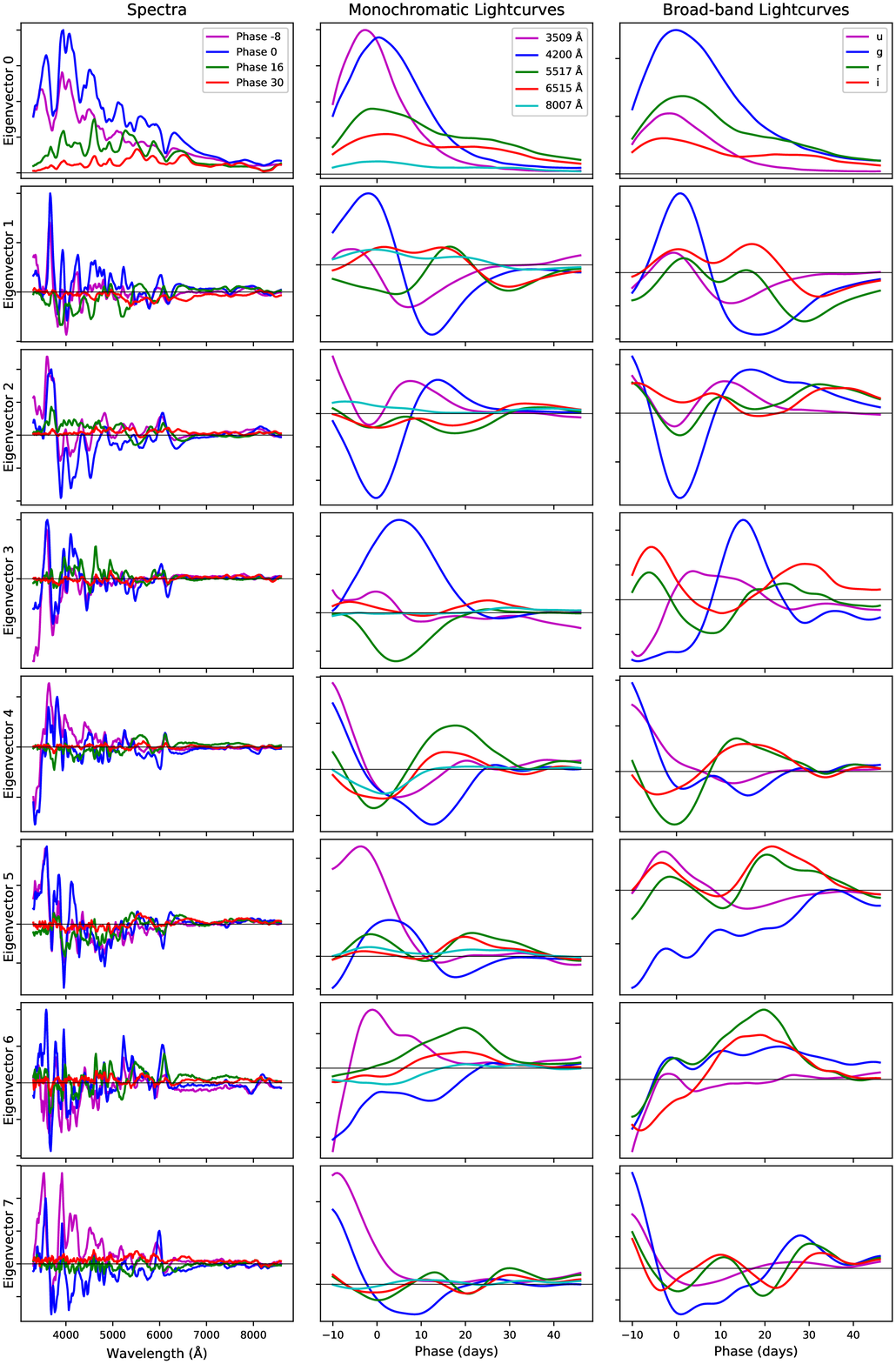}
\centering
\caption{Demonstrative spectra and lightcurves for the first eight eigenvectors of the fourteen-component model trained with no outlier supernovae. See Figure~\ref{fig: Model-2} for description of the columns.}
\label{fig: model 14, no outs, pt1}
\end{figure}

Ultimately it was decided that excluding the peculiar supernovae did not provide enough benefits to outweigh the somewhat subjective decision to exclude them from the model training set. There is a continuum in the behavior that allows these supernovae to be identified as SN~1991T or SN~1991bg-like, and supernovae that share some of these features are seen in our data set that are not outliers in the model coefficient space. This means that to have a model trained only on `normal' supernovae, all of the SN~1991T or SN~1991bg-like supernovae would have to be identified by eye in order to remove them, which would be dependent on some chosen limits and would probably be biased by the phase coverage of the supernovae, since SN~1991T-like supernova spectra appear normal at later phases. On balance, it is preferable to avoid subjective choices in training the model, with the benefit that since the model has been trained on peculiar supernovae it will be easier to identify these supernovae when the model is applied to a new supernova set, and to remove them if necessary for cosmological analyses.

\bibliographystyle{apj}
\bibliography{GP_PCA_bib}

\begin{thebibliography}{}
\expandafter\ifx\csname natexlab\endcsname\relax\def\natexlab#1{#1}\fi

\bibitem[{{Aldering} {et~al.}(2002){Aldering}, {Adam}, {Antilogus}, {Astier},
  {Bacon}, {Bongard}, {Bonnaud}, {Copin}, {Hardin}, {Henault}, {Howell},
  {Lemonnier}, {Levy}, {Loken}, {Nugent}, {Pain}, {Pecontal}, {Pecontal},
  {Perlmutter}, {Quimby}, {Schahmaneche}, {Smadja}, \& {Wood-Vasey}}]{Aldering}
{Aldering}, G., {Adam}, G., {Antilogus}, P., {et~al.} 2002, in \procspie, Vol.
  4836, Survey and Other Telescope Technologies and Discoveries, ed. J.~A.
  {Tyson} \& S.~{Wolff}, 61--72

\bibitem[{{Amanullah} {et~al.}(2010){Amanullah}, {Lidman}, {Rubin}, {Aldering},
  {Astier}, {Barbary}, {Burns}, {Conley}, {Dawson}, {Deustua}, {Doi}, {Fabbro},
  {Faccioli}, {Fakhouri}, {Folatelli}, {Fruchter}, {Furusawa}, {Garavini},
  {Goldhaber}, {Goobar}, {Groom}, {Hook}, {Howell}, {Kashikawa}, {Kim}, {Knop},
  {Kowalski}, {Linder}, {Meyers}, {Morokuma}, {Nobili}, {Nordin}, {Nugent},
  {{\"O}stman}, {Pain}, {Panagia}, {Perlmutter}, {Raux}, {Ruiz-Lapuente},
  {Spadafora}, {Strovink}, {Suzuki}, {Wang}, {Wood-Vasey}, {Yasuda}, \&
  {Supernova Cosmology Project}}]{Amanullah:2010}
{Amanullah}, R., {Lidman}, C., {Rubin}, D., {et~al.} 2010, \apj, 716, 712

\bibitem[{{Bailey} {et~al.}(2009){Bailey}, {Aldering}, {Antilogus}, {Aragon},
  {Baltay}, {Bongard}, {Buton}, {Childress}, {Chotard}, {Copin}, {Gangler},
  {Loken}, {Nugent}, {Pain}, {Pecontal}, {Pereira}, {Perlmutter}, {Rabinowitz},
  {Rigaudier}, {Runge}, {Scalzo}, {Smadja}, {Swift}, {Tao}, {Thomas}, {Wu}, \&
  {Nearby Supernova Factory}}]{Bailey:2009}
{Bailey}, S., {Aldering}, G., {Antilogus}, P., {et~al.} 2009, \aap, 500, L17

\bibitem[{Barbary(2014)}]{Barbary:2014}
Barbary, K. 2014, doi:{10.5281/zenodo.11938}

\bibitem[{{Barone-Nugent} {et~al.}(2012){Barone-Nugent}, {Lidman}, {Wyithe},
  {Mould}, {Howell}, {Hook}, {Sullivan}, {Nugent}, {Arcavi}, {Cenko}, {Cooke},
  {Gal-Yam}, {Hsiao}, {Kasliwal}, {Maguire}, {Ofek}, {Poznanski}, \&
  {Xu}}]{Barone-Nugent:2012}
{Barone-Nugent}, R.~L., {Lidman}, C., {Wyithe}, J.~S.~B., {et~al.} 2012,
  \mnras, 425, 1007

\bibitem[{{Bernstein} {et~al.}(2012){Bernstein}, {Kessler}, {Kuhlmann},
  {Biswas}, {Kovacs}, {Aldering}, {Crane}, {D'Andrea}, {Finley}, {Frieman},
  {Hufford}, {Jarvis}, {Kim}, {Marriner}, {Mukherjee}, {Nichol}, {Nugent},
  {Parkinson}, {Reis}, {Sako}, {Spinka}, \& {Sullivan}}]{Bernstein:2012}
{Bernstein}, J.~P., {Kessler}, R., {Kuhlmann}, S., {et~al.} 2012, \apj, 753,
  152

\bibitem[{{Betoule} {et~al.}(2014){Betoule}, {Kessler}, {Guy}, {Mosher},
  {Hardin}, {Biswas}, {Astier}, {El-Hage}, {Konig}, {Kuhlmann}, {Marriner},
  {Pain}, {Regnault}, {Balland}, {Bassett}, {Brown}, {Campbell}, {Carlberg},
  {Cellier-Holzem}, {Cinabro}, {Conley}, {D'Andrea}, {DePoy}, {Doi}, {Ellis},
  {Fabbro}, {Filippenko}, {Foley}, {Frieman}, {Fouchez}, {Galbany}, {Goobar},
  {Gupta}, {Hill}, {Hlozek}, {Hogan}, {Hook}, {Howell}, {Jha}, {Le Guillou},
  {Leloudas}, {Lidman}, {Marshall}, {M{\"o}ller}, {Mour{\~a}o}, {Neveu},
  {Nichol}, {Olmstead}, {Palanque-Delabrouille}, {Perlmutter}, {Prieto},
  {Pritchet}, {Richmond}, {Riess}, {Ruhlmann-Kleider}, {Sako}, {Schahmaneche},
  {Schneider}, {Smith}, {Sollerman}, {Sullivan}, {Walton}, \&
  {Wheeler}}]{Betoule:2014}
{Betoule}, M., {Kessler}, R., {Guy}, J., {et~al.} 2014, \aap, 568, A22

\bibitem[{{Bohlin}(2014)}]{Bohlin:2014}
{Bohlin}, R.~C. 2014, \aj, 147, 127

\bibitem[{{Bongard} {et~al.}(2011){Bongard}, {Soulez}, {Thi{\'e}baut}, \&
  {Pecontal}}]{Bongard:2011}
{Bongard}, S., {Soulez}, F., {Thi{\'e}baut}, {\'E}., \& {Pecontal}, {\'E}.
  2011, \mnras, 418, 258

\bibitem[{Burnham \& Anderson(2003)}]{burnham2003}
Burnham, K., \& Anderson, D. 2003, Model Selection and Multimodel Inference: A
  Practical Information-Theoretic Approach (Springer New York)

\bibitem[{{Burns} {et~al.}(2011){Burns}, {Stritzinger}, {Phillips}, {Kattner},
  {Persson}, {Madore}, {Freedman}, {Boldt}, {Campillay}, {Contreras},
  {Folatelli}, {Gonzalez}, {Krzeminski}, {Morrell}, {Salgado}, \&
  {Suntzeff}}]{Burns:2010vn}
{Burns}, C.~R., {Stritzinger}, M., {Phillips}, M.~M., {et~al.} 2011, \aj, 141,
  19

\bibitem[{{Buton} {et~al.}(2013){Buton}, {Copin}, {Aldering}, {Antilogus},
  {Aragon}, {Bailey}, {Baltay}, {Bongard}, {Canto}, {Cellier-Holzem},
  {Childress}, {Chotard}, {Fakhouri}, {Gangler}, {Guy}, {Hsiao}, {Kerschhaggl},
  {Kowalski}, {Loken}, {Nugent}, {Paech}, {Pain}, {P{\'e}contal}, {Pereira},
  {Perlmutter}, {Rabinowitz}, {Rigault}, {Runge}, {Scalzo}, {Smadja}, {Tao},
  {Thomas}, {Weaver}, {Wu}, \& {Nearby SuperNova Factory}}]{Buton:2012}
{Buton}, C., {Copin}, Y., {Aldering}, G., {et~al.} 2013, \aap, 549, A8

\bibitem[{{Cardelli} {et~al.}(1989){Cardelli}, {Clayton}, \&
  {Mathis}}]{Cardelli:1989}
{Cardelli}, J.~A., {Clayton}, G.~C., \& {Mathis}, J.~S. 1989, \apj, 345, 245

\bibitem[{{Childress} {et~al.}(2013){Childress}, {Aldering}, {Antilogus},
  {Aragon}, {Bailey}, {Baltay}, {Bongard}, {Buton}, {Canto}, {Cellier-Holzem},
  {Chotard}, {Copin}, {Fakhouri}, {Gangler}, {Guy}, {Hsiao}, {Kerschhaggl},
  {Kim}, {Kowalski}, {Loken}, {Nugent}, {Paech}, {Pain}, {Pecontal}, {Pereira},
  {Perlmutter}, {Rabinowitz}, {Rigault}, {Runge}, {Scalzo}, {Smadja}, {Tao},
  {Thomas}, {Weaver}, \& {Wu}}]{Childress:2013}
{Childress}, M., {Aldering}, G., {Antilogus}, P., {et~al.} 2013, \apj, 770, 107

\bibitem[{{Chotard} {et~al.}(2011){Chotard}, {Gangler}, {Aldering},
  {Antilogus}, {Aragon}, {Bailey}, {Baltay}, {Bongard}, {Buton}, {Canto},
  {Childress}, {Copin}, {Fakhouri}, {Hsiao}, {Kerschhaggl}, {Kowalski},
  {Loken}, {Nugent}, {Paech}, {Pain}, {Pecontal}, {Pereira}, {Perlmutter},
  {Rabinowitz}, {Runge}, {Scalzo}, {Smadja}, {Tao}, {Thomas}, {Weaver}, {Wu},
  \& {Nearby Supernova Factory}}]{Chotard:2011}
{Chotard}, N., {Gangler}, E., {Aldering}, G., {et~al.} 2011, \aap, 529, L4

\bibitem[{{Fakhouri} {et~al.}(2015){Fakhouri}, {Boone}, {Aldering},
  {Antilogus}, {Aragon}, {Bailey}, {Baltay}, {Barbary}, {Baugh}, {Bongard},
  {Buton}, {Chen}, {Childress}, {Chotard}, {Copin}, {Fagrelius}, {Feindt},
  {Fleury}, {Fouchez}, {Gangler}, {Hayden}, {Kim}, {Kowalski}, {Leget},
  {Lombardo}, {Nordin}, {Pain}, {Pecontal}, {Pereira}, {Perlmutter},
  {Rabinowitz}, {Ren}, {Rigault}, {Rubin}, {Runge}, {Saunders}, {Scalzo},
  {Smadja}, {Sofiatti}, {Strovink}, {Suzuki}, {Tao}, {Thomas}, {Weaver}, \&
  {Nearby Supernova Factory}}]{Fakhouri:2015}
{Fakhouri}, H.~K., {Boone}, K., {Aldering}, G., {et~al.} 2015, \apj, 815, 58

\bibitem[{{Filippenko} {et~al.}(1992{\natexlab{a}}){Filippenko}, {Richmond},
  {Matheson}, {Shields}, {Burbidge}, {Cohen}, {Dickinson}, {Malkan}, {Nelson},
  {Pietz}, {Schlegel}, {Schmeer}, {Spinrad}, {Steidel}, {Tran}, \&
  {Wren}}]{Filippenko:1992b}
{Filippenko}, A.~V., {Richmond}, M.~W., {Matheson}, T., {et~al.}
  1992{\natexlab{a}}, \apjl, 384, L15

\bibitem[{{Filippenko} {et~al.}(1992{\natexlab{b}}){Filippenko}, {Richmond},
  {Branch}, {Gaskell}, {Herbst}, {Ford}, {Treffers}, {Matheson}, {Ho}, {Dey},
  {Sargent}, {Small}, \& {van Breugel}}]{Filippenko:1992a}
{Filippenko}, A.~V., {Richmond}, M.~W., {Branch}, D., {et~al.}
  1992{\natexlab{b}}, \aj, 104, 1543

\bibitem[{{Fitzpatrick} \& {Massa}(2007)}]{Fitzpatrick:2007}
{Fitzpatrick}, E.~L., \& {Massa}, D. 2007, \apj, 663, 320

\bibitem[{{Foley} {et~al.}(2018){Foley}, {Hoffmann}, {Macri}, {Riess}, {Brown},
  {Filippenko}, {Graham}, \& {Milne}}]{Foley:2018}
{Foley}, R.~J., {Hoffmann}, S.~L., {Macri}, L.~M., {et~al.} 2018, ArXiv
  e-prints, arXiv:1806.08359

\bibitem[{Ghahramani {et~al.}(1996)Ghahramani, Hinton,
  {et~al.}}]{Ghahramani:1996}
Ghahramani, Z., Hinton, G.~E., {et~al.} 1996, The EM algorithm for mixtures of
  factor analyzers, Tech. rep., Technical Report CRG-TR-96-1, University of
  Toronto

\bibitem[{{Gupta} {et~al.}(2011){Gupta}, {D'Andrea}, {Sako}, {Conroy}, {Smith},
  {Bassett}, {Frieman}, {Garnavich}, {Jha}, {Kessler}, {Lampeitl}, {Marriner},
  {Nichol}, \& {Schneider}}]{Gupta:2011}
{Gupta}, R.~R., {D'Andrea}, C.~B., {Sako}, M., {et~al.} 2011, \apj, 740, 92

\bibitem[{{Guy} {et~al.}(2007){Guy}, {Astier}, {Baumont}, {Hardin}, {Pain},
  {Regnault}, {Basa}, {Carlberg}, {Conley}, {Fabbro}, {Fouchez}, {Hook},
  {Howell}, {Perrett}, {Pritchet}, {Rich}, {Sullivan}, {Antilogus}, {Aubourg},
  {Bazin}, {Bronder}, {Filiol}, {Palanque-Delabrouille}, {Ripoche}, \&
  {Ruhlmann-Kleider}}]{Guy:2007fk}
{Guy}, J., {Astier}, P., {Baumont}, S., {et~al.} 2007, \aap, 466, 11

\bibitem[{{Hamuy} {et~al.}(1994){Hamuy}, {Suntzeff}, {Heathcote}, {Walker},
  {Gigoux}, \& {Phillips}}]{Hamuy:1994}
{Hamuy}, M., {Suntzeff}, N.~B., {Heathcote}, S.~R., {et~al.} 1994, \pasp, 106,
  566

\bibitem[{{Hamuy} {et~al.}(1992){Hamuy}, {Walker}, {Suntzeff}, {Gigoux},
  {Heathcote}, \& {Phillips}}]{Hamuy:1992}
{Hamuy}, M., {Walker}, A.~R., {Suntzeff}, N.~B., {et~al.} 1992, \pasp, 104, 533

\bibitem[{{Hayes} \& {Latham}(1975)}]{Hayes:1975}
{Hayes}, D.~S., \& {Latham}, D.~W. 1975, \apj, 197, 593

\bibitem[{{Howell} {et~al.}(2009){Howell}, {Sullivan}, {Brown}, {Conley}, {Le
  Borgne}, {Hsiao}, {Astier}, {Balam}, {Balland}, {Basa}, {Carlberg},
  {Fouchez}, {Guy}, {Hardin}, {Hook}, {Pain}, {Perrett}, {Pritchet},
  {Regnault}, {Baumont}, {LeDu}, {Lidman}, {Perlmutter}, {Suzuki}, {Walker}, \&
  {Wheeler}}]{Howell:2009}
{Howell}, D.~A., {Sullivan}, M., {Brown}, E.~F., {et~al.} 2009, \apj, 691, 661

\bibitem[{{Hsiao} {et~al.}(2007){Hsiao}, {Conley}, {Howell}, {Sullivan},
  {Pritchet}, {Carlberg}, {Nugent}, \& {Phillips}}]{Hsiao:2007uq}
{Hsiao}, E.~Y., {Conley}, A., {Howell}, D.~A., {et~al.} 2007, \apj, 663, 1187

\bibitem[{{Huang} {et~al.}(2017){Huang}, {Raha}, {Aldering}, {Antilogus},
  {Bailey}, {Baltay}, {Barbary}, {Baugh}, {Boone}, {Bongard}, {Buton}, {Chen},
  {Chotard}, {Copin}, {Fagrelius}, {Fakhouri}, {Feindt}, {Fouchez}, {Gangler},
  {Hayden}, {Hillebrandt}, {Kim}, {Kowalski}, {Leget}, {Lombardo}, {Nordin},
  {Pain}, {Pecontal}, {Pereira}, {Perlmutter}, {Rabinowitz}, {Rigault},
  {Rubin}, {Runge}, {Saunders}, {Smadja}, {Sofiatti}, {Stocker}, {Suzuki},
  {Taubenberger}, {Tao}, {Thomas}, \& {Nearby Supernova Factory}}]{Huang:2017}
{Huang}, X., {Raha}, Z., {Aldering}, G., {et~al.} 2017, \apj, 836, 157

\bibitem[{{Jha} {et~al.}(2007){Jha}, {Riess}, \& {Kirshner}}]{Jha:2007ys}
{Jha}, S., {Riess}, A.~G., \& {Kirshner}, R.~P. 2007, \apj, 659, 122

\bibitem[{{Kelly} {et~al.}(2010){Kelly}, {Hicken}, {Burke}, {Mandel}, \&
  {Kirshner}}]{Kelly:2010}
{Kelly}, P.~L., {Hicken}, M., {Burke}, D.~L., {Mandel}, K.~S., \& {Kirshner},
  R.~P. 2010, \apj, 715, 743

\bibitem[{{Kim} {et~al.}(2013){Kim}, {Thomas}, {Aldering}, {Antilogus},
  {Aragon}, {Bailey}, {Baltay}, {Bongard}, {Buton}, {Canto}, {Cellier-Holzem},
  {Childress}, {Chotard}, {Copin}, {Fakhouri}, {Gangler}, {Guy}, {Kerschhaggl},
  {Kowalski}, {Nordin}, {Nugent}, {Paech}, {Pain}, {Pecontal}, {Pereira},
  {Perlmutter}, {Rabinowitz}, {Rigault}, {Runge}, {Saunders}, {Scalzo},
  {Smadja}, {Tao}, {Weaver}, \& {Wu}}]{Kim:2013}
{Kim}, A.~G., {Thomas}, R.~C., {Aldering}, G., {et~al.} 2013, \apj, 766, 84

\bibitem[{{Krisciunas} {et~al.}(2004){Krisciunas}, {Phillips}, \&
  {Suntzeff}}]{Krisciunas:2004}
{Krisciunas}, K., {Phillips}, M.~M., \& {Suntzeff}, N.~B. 2004, \apjl, 602, L81

\bibitem[{{Kulkarni}(2013)}]{Kulkarni:2013}
{Kulkarni}, S.~R. 2013, The Astronomer's Telegram, 4807

\bibitem[{{Lantz} {et~al.}(2004){Lantz}, {Aldering}, {Antilogus}, {Bonnaud},
  {Capoani}, {Castera}, {Copin}, {Dubet}, {Gangler}, {Henault}, {Lemonnier},
  {Pain}, {Pecontal}, {Pecontal}, \& {Smadja}}]{Lantz:2004}
{Lantz}, B., {Aldering}, G., {Antilogus}, P., {et~al.} 2004, in \procspie, Vol.
  5249, Optical Design and Engineering, ed. L.~{Mazuray}, P.~J. {Rogers}, \&
  R.~{Wartmann}, 146--155

\bibitem[{{LSST Science Collaboration} {et~al.}(2009){LSST Science
  Collaboration}, {Abell}, {Allison}, {Anderson}, {Andrew}, {Angel}, {Armus},
  {Arnett}, {Asztalos}, {Axelrod}, \& et~al.}]{LSST:2009}
{LSST Science Collaboration}, {Abell}, P.~A., {Allison}, J., {et~al.} 2009,
  ArXiv e-prints, arXiv:0912.0201

\bibitem[{{Mahalanobis}(1936)}]{Mahalanobis:1936}
{Mahalanobis}, P.~C. 1936, PNISI, 2, 49

\bibitem[{{Mandel} {et~al.}(2014){Mandel}, {Foley}, \&
  {Kirshner}}]{Mandel:2014}
{Mandel}, K.~S., {Foley}, R.~J., \& {Kirshner}, R.~P. 2014, \apj, 797, 75

\bibitem[{Mandel {et~al.}(2011)Mandel, Narayan, \& Kirshner}]{Mandel:2011zr}
Mandel, K.~S., Narayan, G., \& Kirshner, R.~P. 2011, Astrophys.J., 731, 120

\bibitem[{Mat{\'e}rn(1986)}]{Matern:1986}
Mat{\'e}rn, B. 1986, Spatial variation, Lecture notes in statistics
  (Springer-Verlag)

\bibitem[{{Nordin} {et~al.}(2018){Nordin}, {Aldering}, {Antilogus}, {Aragon},
  {Bailey}, {Baltay}, {Barbary}, {Bongard}, {Boone}, {Brinnel}, {Buton},
  {Childress}, {Chotard}, {Copin}, {Dixon}, {Fagrelius}, {Feindt}, {Fouchez},
  {Gangler}, {Hayden}, {Hillebrandt}, {Kim}, {Kowalski}, {Kuesters}, {Leget},
  {Lombardo}, {Lin}, {Pain}, {Pecontal}, {Pereira}, {Perlmutter}, {Rabinowitz},
  {Rigault}, {Runge}, {Rubin}, {Saunders}, {Smadja}, {Sofiatti}, {Suzuki},
  {Taubenberger}, {Tao}, {Thomas}, \& {Nearby Supernova Factory}}]{Nordin:2018}
{Nordin}, J., {Aldering}, G., {Antilogus}, P., {et~al.} 2018, \aap, 614, A71

\bibitem[{{Nugent} {et~al.}(2002){Nugent}, {Kim}, \&
  {Perlmutter}}]{Nugent:2002fk}
{Nugent}, P., {Kim}, A., \& {Perlmutter}, S. 2002, \pasp, 114, 803

\bibitem[{{Pereira} {et~al.}(2013){Pereira}, {Thomas}, {Aldering}, {Antilogus},
  {Baltay}, {Benitez-Herrera}, {Bongard}, {Buton}, {Canto}, {Cellier-Holzem},
  {Chen}, {Childress}, {Chotard}, {Copin}, {Fakhouri}, {Fink}, {Fouchez},
  {Gangler}, {Guy}, {Hillebrandt}, {Hsiao}, {Kerschhaggl}, {Kowalski},
  {Kromer}, {Nordin}, {Nugent}, {Paech}, {Pain}, {P{\'e}contal}, {Perlmutter},
  {Rabinowitz}, {Rigault}, {Runge}, {Saunders}, {Smadja}, {Tao},
  {Taubenberger}, {Tilquin}, \& {Wu}}]{Pereira:2013}
{Pereira}, R., {Thomas}, R.~C., {Aldering}, G., {et~al.} 2013, \aap, 554, A27

\bibitem[{{Perlmutter} {et~al.}(1999){Perlmutter}, {Aldering}, {Goldhaber},
  {Knop}, {Nugent}, {Castro}, {Deustua}, {Fabbro}, {Goobar}, {Groom}, {Hook},
  {Kim}, {Kim}, {Lee}, {Nunes}, {Pain}, {Pennypacker}, {Quimby}, {Lidman},
  {Ellis}, {Irwin}, {McMahon}, {Ruiz-Lapuente}, {Walton}, {Schaefer}, {Boyle},
  {Filippenko}, {Matheson}, {Fruchter}, {Panagia}, {Newberg}, {Couch}, \&
  {Project}}]{Perlmutter:1999}
{Perlmutter}, S., {Aldering}, G., {Goldhaber}, G., {et~al.} 1999, \apj, 517,
  565

\bibitem[{{Phillips}(1993)}]{Phillips:1993}
{Phillips}, M.~M. 1993, \apjl, 413, L105

\bibitem[{{Riess} {et~al.}(1996){Riess}, {Press}, \& {Kirshner}}]{Riess:1996}
{Riess}, A.~G., {Press}, W.~H., \& {Kirshner}, R.~P. 1996, \apj, 473, 88

\bibitem[{{Riess} {et~al.}(1998){Riess}, {Filippenko}, {Challis},
  {Clocchiatti}, {Diercks}, {Garnavich}, {Gilliland}, {Hogan}, {Jha},
  {Kirshner}, {Leibundgut}, {Phillips}, {Reiss}, {Schmidt}, {Schommer},
  {Smith}, {Spyromilio}, {Stubbs}, {Suntzeff}, \& {Tonry}}]{Riess:1998}
{Riess}, A.~G., {Filippenko}, A.~V., {Challis}, P., {et~al.} 1998, \aj, 116,
  1009

\bibitem[{{Rigault} {et~al.}(2013){Rigault}, {Copin}, {Aldering}, {Antilogus},
  {Aragon}, {Bailey}, {Baltay}, {Bongard}, {Buton}, {Canto}, {Cellier-Holzem},
  {Childress}, {Chotard}, {Fakhouri}, {Feindt}, {Fleury}, {Gangler},
  {Greskovic}, {Guy}, {Kim}, {Kowalski}, {Lombardo}, {Nordin}, {Nugent},
  {Pain}, {P{\'e}contal}, {Pereira}, {Perlmutter}, {Rabinowitz}, {Runge},
  {Saunders}, {Scalzo}, {Smadja}, {Tao}, {Thomas}, \& {Weaver}}]{Rigault:2013}
{Rigault}, M., {Copin}, Y., {Aldering}, G., {et~al.} 2013, \aap, 560, A66

\bibitem[{{Rigault} {et~al.}(2015){Rigault}, {Aldering}, {Kowalski}, {Copin},
  {Antilogus}, {Aragon}, {Bailey}, {Baltay}, {Baugh}, {Bongard}, {Boone},
  {Buton}, {Chen}, {Chotard}, {Fakhouri}, {Feindt}, {Fagrelius}, {Fleury},
  {Fouchez}, {Gangler}, {Hayden}, {Kim}, {Leget}, {Lombardo}, {Nordin}, {Pain},
  {Pecontal}, {Pereira}, {Perlmutter}, {Rabinowitz}, {Runge}, {Rubin},
  {Saunders}, {Smadja}, {Sofiatti}, {Suzuki}, {Tao}, \&
  {Weaver}}]{Rigault:2015}
{Rigault}, M., {Aldering}, G., {Kowalski}, M., {et~al.} 2015, \apj, 802, 20

\bibitem[{{Rigault} {et~al.}(2018){Rigault}, {Brinnel}, {Aldering},
  {Antilogus}, {Aragon}, {Bailey}, {Baltay}, {Barbary}, {Bongard}, {Boone},
  {Buton}, {Childress}, {Chotard}, {Copin}, {Dixon}, {Fagrelius}, {Feindt},
  {Fouchez}, {Gangler}, {Hayden}, {Hillebrandt}, {Howell}, {Kim}, {Kowalski},
  {Kuesters}, {Leget}, {Lombardo}, {Lin}, {Nordin}, {Pain}, {Pecontal},
  {Pereira}, {Perlmutter}, {Rabinowitz}, {Runge}, {Rubin}, {Saunders},
  {Smadja}, {Sofiatti}, {Suzuki}, {Taubenberger}, {Tao}, \&
  {Thomas}}]{Rigault:2018}
{Rigault}, M., {Brinnel}, V., {Aldering}, G., {et~al.} 2018, ArXiv e-prints,
  arXiv:1806.03849

\bibitem[{{Sasdelli} {et~al.}(2015){Sasdelli}, {Hillebrandt}, {Aldering},
  {Antilogus}, {Aragon}, {Bailey}, {Baltay}, {Benitez-Herrera}, {Bongard},
  {Buton}, {Canto}, {Cellier-Holzem}, {Chen}, {Childress}, {Chotard}, {Copin},
  {Fakhouri}, {Feindt}, {Fink}, {Fleury}, {Fouchez}, {Gangler}, {Guy},
  {Ishida}, {Kim}, {Kowalski}, {Kromer}, {Lombardo}, {Mazzali}, {Nordin},
  {Pain}, {P{\'e}contal}, {Pereira}, {Perlmutter}, {Rabinowitz}, {Rigault},
  {Runge}, {Saunders}, {Scalzo}, {Smadja}, {Suzuki}, {Tao}, {Taubenberger},
  {Thomas}, {Tilquin}, \& {Weaver}}]{Sasdelli:2015}
{Sasdelli}, M., {Hillebrandt}, W., {Aldering}, G., {et~al.} 2015, \mnras, 447,
  1247

\bibitem[{{Saunders} {et~al.}(2015){Saunders}, {Aldering}, {Antilogus},
  {Aragon}, {Bailey}, {Baltay}, {Bongard}, {Buton}, {Canto}, {Cellier-Holzem},
  {Childress}, {Chotard}, {Copin}, {Fakhouri}, {Feindt}, {Gangler}, {Guy},
  {Kerschhaggl}, {Kim}, {Kowalski}, {Nordin}, {Nugent}, {Paech}, {Pain},
  {Pecontal}, {Pereira}, {Perlmutter}, {Rabinowitz}, {Rigault}, {Rubin},
  {Runge}, {Scalzo}, {Smadja}, {Tao}, {Thomas}, {Weaver}, {Wu}, \& {Nearby
  Supernova Factory}}]{Saunders:2015}
{Saunders}, C., {Aldering}, G., {Antilogus}, P., {et~al.} 2015, \apj, 800, 57

\bibitem[{{Scalzo} {et~al.}(2012){Scalzo}, {Aldering}, {Antilogus}, {Aragon},
  {Bailey}, {Baltay}, {Bongard}, {Buton}, {Canto}, {Cellier-Holzem},
  {Childress}, {Chotard}, {Copin}, {Fakhouri}, {Gangler}, {Guy}, {Hsiao},
  {Kerschhaggl}, {Kowalski}, {Nugent}, {Paech}, {Pain}, {Pecontal}, {Pereira},
  {Perlmutter}, {Rabinowitz}, {Rigault}, {Runge}, {Smadja}, {Tao}, {Thomas},
  {Weaver}, {Wu}, \& {Nearby Supernova Factory}}]{Scalzo:2012}
{Scalzo}, R., {Aldering}, G., {Antilogus}, P., {et~al.} 2012, \apj, 757, 12

\bibitem[{{Scalzo} {et~al.}(2014){Scalzo}, {Aldering}, {Antilogus}, {Aragon},
  {Bailey}, {Baltay}, {Bongard}, {Buton}, {Cellier-Holzem}, {Childress},
  {Chotard}, {Copin}, {Fakhouri}, {Gangler}, {Guy}, {Kim}, {Kowalski},
  {Kromer}, {Nordin}, {Nugent}, {Paech}, {Pain}, {Pecontal}, {Pereira},
  {Perlmutter}, {Rabinowitz}, {Rigault}, {Runge}, {Saunders}, {Sim}, {Smadja},
  {Tao}, {Taubenberger}, {Thomas}, {Weaver}, \& {Nearby Supernova
  Factory}}]{Scalzo:2014}
---. 2014, \mnras, 440, 1498

\bibitem[{{Scalzo} {et~al.}(2010){Scalzo}, {Aldering}, {Antilogus}, {Aragon},
  {Bailey}, {Baltay}, {Bongard}, {Buton}, {Childress}, {Chotard}, {Copin},
  {Fakhouri}, {Gal-Yam}, {Gangler}, {Hoyer}, {Kasliwal}, {Loken}, {Nugent},
  {Pain}, {P{\'e}contal}, {Pereira}, {Perlmutter}, {Rabinowitz}, {Rau},
  {Rigaudier}, {Runge}, {Smadja}, {Tao}, {Thomas}, {Weaver}, \&
  {Wu}}]{Scalzo:2010}
{Scalzo}, R.~A., {Aldering}, G., {Antilogus}, P., {et~al.} 2010, \apj, 713,
  1073

\bibitem[{{Schlafly} {et~al.}(2010){Schlafly}, {Finkbeiner}, {Schlegel},
  {Juri{\'c}}, {Ivezi{\'c}}, {Gibson}, {Knapp}, \& {Weaver}}]{Schlafly:2010}
{Schlafly}, E.~F., {Finkbeiner}, D.~P., {Schlegel}, D.~J., {et~al.} 2010, \apj,
  725, 1175

\bibitem[{{Schlegel} {et~al.}(1998){Schlegel}, {Finkbeiner}, \&
  {Davis}}]{Schlegel:1998}
{Schlegel}, D.~J., {Finkbeiner}, D.~P., \& {Davis}, M. 1998, \apj, 500, 525

\bibitem[{{Sullivan} {et~al.}(2010){Sullivan}, {Conley}, {Howell}, {Neill},
  {Astier}, {Balland}, {Basa}, {Carlberg}, {Fouchez}, {Guy}, {Hardin}, {Hook},
  {Pain}, {Palanque-Delabrouille}, {Perrett}, {Pritchet}, {Regnault}, {Rich},
  {Ruhlmann-Kleider}, {Baumont}, {Hsiao}, {Kronborg}, {Lidman}, {Perlmutter},
  \& {Walker}}]{Sullivan:2010}
{Sullivan}, M., {Conley}, A., {Howell}, D.~A., {et~al.} 2010, \mnras, 406, 782

\bibitem[{{Thomas} {et~al.}(2007){Thomas}, {Aldering}, {Antilogus}, {Aragon},
  {Bailey}, {Baltay}, {Baron}, {Bauer}, {Buton}, {Bongard}, {Copin}, {Gangler},
  {Gilles}, {Kessler}, {Loken}, {Nugent}, {Pain}, {Parrent}, {P{\'e}contal},
  {Pereira}, {Perlmutter}, {Rabinowitz}, {Rigaudier}, {Runge}, {Scalzo},
  {Smadja}, {Wang}, {Weaver}, \& {Nearby Supernova Factory}}]{Thomas:2007}
{Thomas}, R.~C., {Aldering}, G., {Antilogus}, P., {et~al.} 2007, \apjl, 654,
  L53

\bibitem[{{Thomas} {et~al.}(2011){Thomas}, {Aldering}, {Antilogus}, {Aragon},
  {Bailey}, {Baltay}, {Bongard}, {Buton}, {Canto}, {Childress}, {Chotard},
  {Copin}, {Fakhouri}, {Gangler}, {Hsiao}, {Kerschhaggl}, {Kowalski}, {Loken},
  {Nugent}, {Paech}, {Pain}, {Pecontal}, {Pereira}, {Perlmutter}, {Rabinowitz},
  {Rigault}, {Rubin}, {Runge}, {Scalzo}, {Smadja}, {Tao}, {Weaver}, {Wu},
  {Brown}, {Milne}, \& {Nearby Supernova Factory}}]{Thomas:2011}
---. 2011, \apj, 743, 27

\bibitem[{{Tripp}(1998)}]{Tripp:1998}
{Tripp}, R. 1998, \aap, 331, 815

\end{thebibliography}

\end{document}